\documentclass[11pt,a4paper,oneside]{book}

\usepackage{geometry}
\usepackage[caption = false]{subfig}
\usepackage{xcolor,colortbl}
\usepackage{floatrow}
\usepackage{chngcntr}

\usepackage{framed}
\usepackage{xcolor}
\colorlet{shadecolor}{gray!18}

\usepackage{mdframed}
\usepackage{url}
\usepackage{endnotes}
\usepackage{paralist}

\usepackage[utf8]{inputenc}

\usepackage{amsmath}
\usepackage{amssymb,amsthm}
\usepackage{amsfonts,color}
\usepackage{multirow}
\usepackage{array}
\usepackage{enumerate}
\usepackage{mathtools}
\usepackage[final]{pdfpages}\includepdfset{pages=-,noautoscale}
\usepackage{tikz}
\usepackage{tikz-3dplot} 
\usepackage{pgfplots}
\pgfplotsset{compat=1.13}
\usepackage{graphicx}
\usepackage{algorithmic}
\usepackage[toc,page]{appendix}
\usepackage{booktabs}
\usepackage{longtable}
\usepackage{floatrow}
\usepackage{hyperref}
\definecolor{navyblue}{rgb}{0.0, 0.0, 0.5}
\hypersetup{
    colorlinks,
    citecolor=navyblue,
    filecolor=navyblue,
    linkcolor=navyblue,
    urlcolor=navyblue
}

\usepackage[nocompress]{cite}

\newcommand{\idHCI}{\boxed{P1}}
\newcommand{\idTISMIR}{\boxed{P2}}
\newcommand{\idIUI}{\boxed{P3}}
\newcommand{\idFNT}{\boxed{P4}}
\newcommand{\idRecSysACTR}{\boxed{P5}}
\newcommand{\idRecSysTRUST}{\boxed{P6}}
\newcommand{\idUMUAI}{\boxed{P7}}
\newcommand{\idEcirMETA}{\boxed{P8}}
\newcommand{\idTIST}{\boxed{P9}}
\newcommand{\idFRONTPRI}{\boxed{P10}}
\newcommand{\idEcirPOP}{\boxed{P11}}
\newcommand{\idEPJ}{\boxed{P12}}
\newcommand{\idRecSysLBR}{\boxed{P13}}
\newcommand{\idBiasMEDIA}{\boxed{P14}}
\newcommand{\idEcirPRESSE}{\boxed{P15}}
\newcommand{\idBiasCALIBRATION}{\boxed{P16}}
\newcommand{\idSCIREP}{\boxed{P17}}

\newcounter{findingscounter}
\newcommand{\findingbox}[1]{%
    \stepcounter{findingscounter}%
    \vspace{-1ex} %
    \vspace{2mm}
    \begin{center} %
        \noindent %
        \framebox{\parbox{.97\columnwidth}{%
            \textbf{Summary of own research (\thefindingscounter):} #1}}%
    \end{center} %
    \vspace{-1ex}}

\def\blankpage{%
      \clearpage%
      \thispagestyle{empty}%
      \addtocounter{page}{-1}%
      \null%
      \clearpage}

\DeclareMathOperator*{\argmax}{arg\,max}

\newcommand*{\titleGP}{\begingroup
\centering
\vspace*{\baselineskip}

\rule{\textwidth}{1.6pt}\vspace*{-\baselineskip}\vspace*{2pt} 
\rule{\textwidth}{0.4pt}\\[\baselineskip] 
{\LARGE Transparency, Privacy, and Fairness\\in\\\vspace{3mm}Recommender Systems}\\[0.2\baselineskip] 
\rule{\textwidth}{0.4pt}\vspace*{-\baselineskip}\vspace{3.2pt} 
\rule{\textwidth}{1.6pt}\\[\baselineskip] 

\scshape 
Cumulative Habilitation\\for the scientific subject\\[\baselineskip]
APPLIED COMPUTER SCIENCE\par

\vspace*{2\baselineskip} 

Submitted by \\[\baselineskip]
{\Large Dr. Dominik Kowald}\\ [\baselineskip]
{\itshape Institute of Interactive Systems and Data Science\\
  Graz University of Technology\par}

\vfill

\begin{center}
\includegraphics{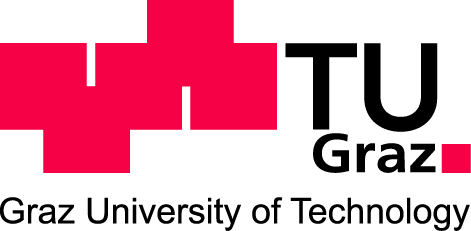} \\[4\baselineskip]
Submitted: October 2023\\
Accepted: June 2024
\end{center}
\endgroup}

\begin{document}

\titleGP
\thispagestyle{empty}

\blankpage{}
\pagestyle{plain}

\newpage

\pagenumbering{roman}

\begin{center}
  \vspace*{5cm}
  
  \textit{I've got to keep going, be strong. Must be so determined and push myself on.}\\ \vspace*{5mm}--- Iron Maiden, The Loneliness of the Long Distance Runner ---\\

  \vspace*{7cm}

  Dedicated to my wife, Tea.

\end{center}



\chapter*{Overview of Chapters}
\subsubsection{Chapter~\ref{c:intro} - Introduction}
This chapter starts with the motivation and the scientific positioning of this habilitation within the broad research field of recommender systems. It also lists and briefly outlines the 17 main publications of this habilitation.

\subsubsection{Chapter~\ref{c:relwork} - Related Work and Background}
Chapter 2 briefly discusses the related work and background relevant to this habilitation, namely (i) main concepts of recommender systems, (ii) transparency and cognitive models in recommender systems, (iii) privacy and limited preference information in recommender systems, and (iv) fairness and popularity bias in recommender systems. Additionally, this chapter briefly summarizes the author's own research efforts in relation to the related work.

\subsubsection{Chapter~\ref{c:contributions} - Scientific Contributions}
This chapter describes the 7 scientific contributions of this habilitation, which are (i) using cognitive models for a transparent design and implementation process of recommender systems, (ii) illustrating to what extent components of the cognitive model ACT-R contribute to recommendations, (iii) addressing limited user preference information in cold-start and session-based recommendation settings, (iv) addressing users' privacy constraints and the trade-off between accuracy and privacy in recommendations, (v) measuring popularity bias for user groups differing in mainstreaminess and gender, (vi) understanding popularity bias mitigation and amplification, and (vii) studying long-term dynamics of fairness in algorithmic decision support. Additionally, this chapter discusses reproducibility aspects of the presented research results and findings.

\subsubsection{Chapter~\ref{c:outlook} - Outlook and Future Research}
Chapter 4 gives an outlook into future research directions based on the results, scientific contributions, and findings of this habilitation.

\vspace{8mm} \noindent \textbf{Please note} that this is a slightly adapted and updated version of this habilitation reflecting the state of the research conducted  until mid 2024. Furthermore, this version of the habilitation does not contain the full texts of the publications, but the DOI links to the online versions are provided in Section~\ref{sec:main_pubs}. 



\tableofcontents



\chapter*{Acknowledgements}
\addcontentsline{toc}{chapter}{Acknowledgements} 

I would like to thank a number of excellent people who supported me in the last six years, while working on his habilitation. First, Stefanie Lindstaedt, the former head of the Institute of Interactive Systems and Data Science (ISDS) of TU Graz, and former CEO of Know-Center Graz, for supporting me during my whole habilitation process, and in building up my own group, the FAIR-AI research area. I wish you all the best for your next endeavors, namely being the founding president of the Institute of Digital Sciences Austria (IDSA) in Linz. I also thank Frank Kappe, who is the new head of ISDS, and Roman Kern, who is taking over the scientific leadership of Know-Center Graz, for supporting me in my final steps of this habilitation process. Another special thank you goes to Elisabeth Lex, my Ph.D. mentor and former area head of the Social Computing group at Know-Center Graz. Thank you for all the valuable advices with respect to this habilitation, and for the great collaborations on our joint publications.

Within ISDS and Know-Center Graz, I would like to thank my FAIR-AI research group for the great support, and for jointly working together on all our research projects. Special thanks go to Emanuel Lacic, now working at InfoBip in Zagreb, for helping me in building up this research group, to Simone Kopeinik for bringing in all these excellent research ideas with respect to fairness and bias in AI, to Dieter Theiler for providing important software development know-how needed to deploy recommender systems in practice, to Leon Fadljevic for the support in all our data science projects, to Peter Muellner for being an excellent PhD student, and for always finishing the planned tasks perfectly in time, and, to Tomislav Duricic for being another great PhD student, and for always being highly motivated independent of the given task. Finally, I also thank Jana Lasser for providing me with valuable advices for finishing this habilitation.

I have been lucky to collaborate with a lot of brilliant people in the last years: I would like to thank Markus Schedl from Johannes Kepler University in Linz for the fruitful collaborations and research projects we conducted so far, Nava Tintarev from Maastricht University for being a great host during my research visit in 2021, Robin Burke from University Boulder-Colorado for the interesting discussions on fairness aspects of recommender systems, and Eva Zangerle from University Innsbruck and Christine Bauer from Paris-Lodron University Salzburg for the great collaborations. I am looking forward to our discussions at the Dagstuhl Seminar in 2024! I also want to thank the members of the commission and the reviewers of this habilitation for their work and valuable feedback. Finally, I thank my friends and family, and here especially my wife Tea, for their support, understanding, and love. Rest in Peace, Philipp, we will never forget you!


\chapter*{Abstract}
\addcontentsline{toc}{chapter}{Abstract}

Recommender systems have become a pervasive part of our daily online experience by analyzing past usage behavior to suggest potential relevant content, e.g., music, movies, or books. Today, recommender systems are one of the most widely used applications of artificial intelligence and machine learning. Therefore, regulations and requirements for trustworthy artificial intelligence, for example, the European AI Act, which includes notions such as transparency, privacy, and fairness are also highly relevant for the design, development, evaluation, and deployment of recommender systems in practice. This habilitation elaborates on aspects related to these three notions in the light of recommender systems, namely: (i) transparency and cognitive models, (ii) privacy and limited preference information, and (iii) fairness and popularity bias in recommender systems. 

Specifically, with respect to aspect (i), we highlight the usefulness of incorporating psychological theories for a transparent design process of recommender systems. We term this type of systems psychology-informed recommender systems. We also use models of human memory theory to develop cognitive-inspired algorithms for tag and music recommendations, and find that these algorithms are capable of outperforming related methods in terms of recommendation accuracy. Additionally, we show that cognitive models can further contribute to transparency aspects of recommender systems by illustrating how the models' components have contributed to generate the recommendation lists.

In aspect (ii), we study and address the trade-off between accuracy and privacy in differentially-private recommendations. We design a novel recommendation approach for collaborative filtering based on an efficient neighborhood reuse concept, which reduces the number of users that need to be protected with differential privacy. Furthermore, we address the related issue of limited availability of user preference information, e.g., click data, in the settings of session-based and cold-start recommendations, by using, e.g., variational autoencoders.

With respect to aspect (iii), we analyze popularity bias in collaborative filtering-based recommender systems. We find that the recommendation frequency of an item is positively correlated with this item's popularity. This also leads to the unfair treatment of users with little interest in popular content, since these users receive worse recommendation accuracy results than users with high interest in popular content. We also find that female users are more strongly affected by the algorithms' amplification of popularity bias. Besides, we present results of an online study on popularity bias mitigation in the field of news article recommendations. Finally, we study long-term fairness dynamics in algorithmic decision support in the labor market using agent-based modeling techniques.


\chapter*{Kurzfassung}
\addcontentsline{toc}{chapter}{Kurzfassung (Abstract in German)}

Empfehlungssysteme sind zu einem allgegenwärtigen Teil unserer täglichen Online-Erfahrung geworden, indem sie das vergangene Nutzerverhalten analysieren, um relevante Inhalte vorzuschlagen, beispielsweise Musik, Filme oder Bücher. Mittlerweile gehören Empfehlungssysteme zu den am weitesten verbreiteten Anwendungen der künstlichen Intelligenz und des maschinellen Lernens. Daher sind Vorschriften für vertrauenswürdige künstliche Intelligenz, welche Anforderungen wie Transparenz, Datenschutz und Fairness umfassen, für die Entwicklung von Empfehlungssystemen relevant. Diese Habilitation untersucht Empfehlungssysteme in Hinblick auf Aspekte, die mit diesen Anforderungen verknüpft sind, nämlich: (i) Transparenz und kognitive Modelle, (ii) Datenschutz und limitierte Präferenz-Informationen, sowie (iii) Fairness und Popularitätsverzerrungen.

Bezüglich Aspekt (i) zeigen wir den Nutzen von psychologischen Theorien für einen transparenten Designprozess von Empfehlungssystemen. Wir bezeichnen diese als Psychologie-inspirierte Empfehlungssysteme. Zusätzlich verwenden wir Modelle der menschlichen Gedächtnistheorie für die Entwicklung von Empfehlungssystemen und zeigen, dass diese Algorithmen verwandte Methoden, in Bezug auf die Vorhersagegenauigkeit, übertreffen. Darüber hinaus zeigen wir, dass die kognitiven Modelle dazu verwendet werden können, um zu illustrieren, welche Komponenten für die Empfehlungsgenerierung wichtig gewesen sind.

In Hinblick auf Aspekt (ii) untersuchen wir die Beziehung zwischen Genauigkeit und Datenschutz in Empfehlungssystemen, die Differential Privacy verwenden. Wir entwickeln einen neuartigen Empfehlungsalgorithmus, der auf einem effizienten Konzept zur Wiederverwendung von Nachbarschaften im kollaborativen Filtern basiert. Dadurch kann der notwendige Einsatz von Differential Privacy minimiert werden. Darüber hinaus adressieren wir ein damit verwandtes Problem, nämlich das der limitierten Nutzerpräferenz-Informationen, z.B., Klick-Daten, durch die Verwendung von z.B., Variational Autoencodern. 

Bezüglich Aspekt (iii) analysieren wir den Einfluss der Popularitätsverzerrung auf die Genauigkeit von Empfehlungssystemen. Wir zeigen, dass Popularität und Empfehlungshäufigkeit positiv korreliert sind, welches auch zur unfairen Behandlung von Nutzern führt, die wenig Interesse an populären Inhalten haben. Diese Nutzer erhalten eine geringere Empfehlungsgenauigkeit als Nutzer, die an populären Inhalten interessiert sind. Darüber hinaus zeigen wir, dass weibliche Benutzer stärker von Popularitätsverzerrungen betroffen sind. Wir präsentieren außerdem Ergebnisse einer Online-Studie zur Minderung des Einflusses von Popularitätsverzerrungen. Abschließend untersuchen wir Langzeiteffekte von Fairness in algorithmischen Entscheidungen mittels agentenbasierter Modellierung.


\blankpage{}
\newpage

\pagenumbering{arabic}

\chapter{Introduction}
\label{c:intro}

The present postdoctoral thesis is a cumulative habilitation submitted to Graz University of Technology for the scientific subject \textit{Applied Computer Science}. This habilitation summarizes and discusses scientific publications that have been published between 2018 and 2023, i.e., during the habilitation's author's postdoctoral research. This chapter describes the scientific positioning of this habilitation (Section~\ref{s:intro_motivation}), and introduces the 17 main publications that constitute this work (Section~\ref{sec:main_pubs}). All publications are peer-reviewed, are already published, and contain a digital object identifier (\textit{DOI}). The publications consist of 7 journal articles, 7 conference proceedings contributions, two workshop post-proceedings book chapters, and one workshop paper. The latter was published via the academic distribution service \textit{arXiv} in accordance with the publishing guidelines of the \textit{Workshop on Transparency and Explainability in Adaptive Systems through User Modeling Grounded in Psychological Theory} co-located with \textit{ACM IUI 2020}.

\section{Scientific Positioning of this Habilitation}
\label{s:intro_motivation}

This habilitation investigates the research field of recommender systems in general, and aspects of transparency and cognitive models, privacy and limited preference information, and fairness and popularity bias in recommender systems in particular. The research field of recommender systems makes use of multiple aspects of \textit{Applied Computer Science}, including (but not limited to) data science, user modeling, personalization, machine learning, information retrieval, human computer interaction, computational social science, and trustworthy artificial intelligence.

More concretely, recommender systems can be seen as one of the most widely used instantiations of machine learning and artificial intelligence, and accompany us in our daily online experience. Thus, recommender systems have become an integral part of our digital life for supporting humans in finding relevant information in information spaces that are too big or complex for manual filtering (e.g.,~\cite{ricci2010introduction,burke2011recommender,jannach2016recommender,frontiers_editorial_2024}). Since the early implementations of recommendation algorithms (e.g.,~\cite{resnick1994grouplens,resnick1997recommender}), these systems analyze past usage behavior in order to build user models, and to suggest items, or even people in social networks~\cite{kowald2013social,lacic2015utilizing,eirinaki2018recommender}, to individual users or to groups of users (e.g.,~\cite{masthoff2010group,masthoff2022group}). To build these user models, different techniques have been employed, including traditional approaches such as collaborative filtering~\cite{ekstrand2011collaborative}, content-based filtering~\cite{lops2010content}, and hybrid recommendations~\cite{burke2002hybrid}, and more recent approaches based on latent representations (or embeddings) and deep learning~\cite{wang2015collaborative,zhang2019deep,chen2023deep}. Thus, also different types of data sources are utilized for generating recommendations, e.g., preference information such as ratings, and content features of items (see Section~\ref{s:rel_main} for more details on recommender systems in general). 
Apart from that, recent research has illustrated the multi-stakeholder nature of recommender systems~\cite{abdollahpouri2017recommender,abdollahpouri2020multistakeholder}. Thus, not only users are affected by recommendations, but also other stakeholders~\cite{jannach2020escaping}, such as platform operators or item providers (e.g., music artists). Balancing the goals of multiple stakeholders is an active research topic, and further illustrates the far-reaching impact of recommender systems on society~\cite{burke2018balanced}.

The uptake of recommender systems both in academia and industry~\cite{jannach2022impact,jannach2016recommendations,castells2023recommender}, as well as their human-centric nature, emphasizes that current regulations and requirements for trustworthy artificial intelligence (AI) are also of high importance for the deployment of recommender systems~\cite{di2022recommender}. Trustworthiness entails multiple notions that have been defined and categorized by the European Commission and institutions in other countries. This has led to different regulations and requirements, for example, the \textit{EU Artificial Intelligence Act}~\cite{webaiact}, or the \textit{United States Regulatory Development Relating to AI}~\cite{webusact}. Although there are differences between these regulations and requirements, all of them include notions related to transparency, privacy, and fairness in AI. These aspects are also highly relevant under the lens of recommender systems, as indicated by recent related research investigating trustworthy recommender systems~\cite{fan2022comprehensive,ge2022survey,fan2023trustworthy}. This habilitation contributes to this line of research in the following fields: 

\subsubsection{Transparency and Cognitive Models in Recommender Systems}

One issue of modern recommender systems algorithms based on deep learning techniques (e.g.,~\cite{wang2015collaborative,zhang2019deep}) is that these approaches are mostly based on principles of artificial intelligence rather than human intelligence. This could lead to non-transparent algorithmic decisions that are hard to understand by the system's users~\cite{sinha2002role}. Apart from methods coming from the fields of explainable AI~\cite{miller2019explanation} and explainable recommender systems~\cite{tintarev2015explaining}, one way to address this issue is to use theories from psychology to enhance the transparency of recommendation models.

This habilitation uses cognitive models of human memory for a transparent design of recommendation approaches~\cite{seitlinger2018balancing,kowald2018impact,lex2020modeling,kowald2020utilizing,humanizebook}. Specifically, we show that models of human episodic memory and activation processes in human memory can help to create transparent and accurate recommendation models. In this respect, we also illustrate to what extent the components of these models contribute to the generation of the recommendation lists~\cite{recsys_actr_2023}. Finally, we survey and categorize research at the intersection of recommender systems and psychology, which we term \textit{psychology-informed recommender systems}~\cite{lex2021psychology}.

\subsubsection{Privacy and Limited Preference Information in Recommender Systems}

Recommender systems need to analyze user preference information to calculate personalized recommendations, which could lead to multiple privacy threats to users~\cite{friedman2015privacy}. This includes the inference of users' sensitive information (e.g., gender), or the disclosure of users' preference information (e.g., who bought what) via the analysis of generated recommendation lists by untrusted third parties (e.g.,~\cite{calandrino2011you,beigi2020survey,zhang2023comprehensive}). Thus, privacy has become a key requirement for personalized recommender systems, especially in the light of current data protection initiatives such as the \textit{European General Data Protection Regulation (GDPR)}. Therefore, privacy is related to the issue of limited availability of user preference information (e.g., clicks or ratings) due to the restricted utilization of users' preference information as a result of data protection initiatives~\cite{cummings2018role,biega2020operationalizing}, and due to the increased privacy concerns of users (e.g., users are not willing to share preference information or to sign in to the system)~\cite{knijnenburg2013making,larson2017towards,mehdy2021privacy}. This could lead to the user cold-start problem~\cite{schein2002methods} and session-based recommendation settings, since long-term user preferences (including past preferences of the target user) are unavailable~\cite{jannach2022session}. 

This habilitation investigates issues of limited preference information by addressing the user cold-start problem using recommendations based on users' trust connections~\cite{duricic2018trust}, and by studying the usefulness of variational autoencoders for session-based job recommendations~\cite{lacic2020using}. Additionally, we study varying privacy constraints of users in a matrix factorization-based recommender system using \textit{meta learning}~\cite{muellner2021robustness}. We also address the privacy-accuracy trade-off in differentially private recommender systems by utilizing an efficient neighborhood reuse concept~\cite{tist_dp_2023}. Finally, we survey and categorize the literature on employing \textit{differential privacy} for collaborative filtering recommender systems~\cite{frontiers_privacy_2023}. 

\subsubsection{Fairness and Popularity Bias in Recommender Systems}

Although bias and fairness in algorithmic decision support and machine learning is a research topic that has gained a lot of attraction in recent years~\cite{kusner2017counterfactual,berk2018fairness,mehrabi2021survey}, the reflection and replication of biases is still an open research problem in the field of interactive systems in general~\cite{friedman1996bias,lambrecht2019algorithmic}, and recommender systems in particular~\cite{mansoury2020feedback,melchiorre2020personality,chen2023bias}. Here, especially popularity bias is a common issue in recommender systems based on collaborative filtering, and leads to the underrepresentation of unpopular content in personalized recommendation lists~\cite{elahi2021investigating,abdollahpouri2021user,ahanger2022popularity}.

The research presented in this habilitation shows that this popularity bias unfairly affects users with little interest in popular content, since this user group receives lower recommendation accuracy than users interested in popular content~\cite{kowald2020unfairness,kowald2021support,kowald2022popularity,ecir_bias_2023}. We also find that recommendation algorithms could amplify popularity bias for female users~\cite{lesota2021analyzing}, and that content-based recommendations can help to mitigate popularity bias~\cite{lacic2022drives}. Additionally, we study long-term fairness in algorithmic decision support in the labor market, and find that there is a trade-off between \textit{individual} and \textit{group fairness} in this setting~\cite{scher2023modelling}.

\subsubsection{Reproducibility Aspects of this Habilitation}
The reproducibility of recommender systems research results is of utmost importance to be able to track the scientific progress in the field (e.g.,~\cite{beel2016towards,ferrari2021troubling}). This habilitation provides code and data resources that should foster the reproducibility of the presented research contributions (see Section~\ref{sec:cont_summary} for a full list).

\section{Main Publications}
\label{sec:main_pubs}

Table~\ref{tab:papers} lists the 17 main publications of this habilitation. I have selected 5 publications for each of the first two research topics described beforehand. For the third research topic, fairness and popularity bias in recommender systems, I have selected 7 publications, since this is the research topic I have investigated most recently (here, my first paper was published in 2020). Within these three research fields, the publications are sorted by publication year in ascending order. Overall, each publication is assigned a unique ID, i.e.,~\boxed{$Pi$}, where $i = 1 \dots 17$. 

In the first field, transparency and cognitive models in recommender systems, the list of publications contains three studies, in which cognitive models are employed for a transparent design process of recommender systems, i.e., one recommendation approach based on a model of human episodic memory~\idHCI, and two approaches based on models formalizing activation processes in human memory~\idTISMIR~\idIUI. Furthermore, it lists a survey on psychology-informed recommender systems~\idFNT. Another publication illustrates to what extent components of cognitive models contribute to the generation of the recommendation lists~\idRecSysACTR. 

The second research field contains two studies on addressing the issue of limited availability of user preference information: one addresses the user cold-start problem using trust-based collaborative filtering~\idRecSysTRUST, and one employs variational autoencoders for session-based job recommendations~\idUMUAI. Table~\ref{tab:papers} also contains three publications on privacy-aware recommender systems, one addressing varying privacy constraints of users~\idEcirMETA, one addressing the accuracy-privacy trade-off of differentially private recommender systems~\idTIST, and one surveying the use of \textit{differential privacy} in collaborative filtering recommender systems~\idFRONTPRI. 

In the third field, Table~\ref{tab:papers} contains two publications that study popularity bias and characteristics of ``niche'' users in music recommendations~\idEcirPOP~\idEPJ. One paper further studies if users of different genders are equally affected by popularity bias in music recommendations~\idRecSysLBR, and another paper studies popularity bias in multimedia recommendation domains~\idBiasMEDIA. Furthermore, this list contains an online study on popularity bias mitigation in news article recommendations~\idEcirPRESSE. Another paper analyzes miscalibration and popularity bias amplification in recommendations~\idBiasCALIBRATION. Finally, one journal article studies long-term dynamics of fairness in algorithmic decision support in the labor market~\idSCIREP.  

\begin{longtable}{p{1cm} p{12cm}}
\caption{List of main publications selected by the author of this habilitation.} \\\hline

\textbf{No.} & \textbf{Publication}\\\hline\hline

& \textbf{Transparency and Cognitive Models in Recommender Systems}\\\hline

\idHCI & Seitlinger, P., Ley, T., \textbf{Kowald, D.}, Theiler, D., Hasani-Mavriqi, I., Dennerlein, S., Lex, E., Albert, D. (2018). Balancing the Fluency-Consistency Tradeoff in Collaborative Information Search with a Recommender Approach. 
\textit{International Journal of Human–Computer Interaction}, 34:6, pp. 557-575.   
DOI: \url{https://doi.org/10.1080/10447318.2017.1379240}\\\hline
  
\idTISMIR & Lex, E.*, \textbf{Kowald, D.*},  Schedl, M. (2020). Modeling Popularity and Temporal Drift of Music Genre Preferences. \textit{Transactions of the International Society for Music Information Retrieval}, 3:1, pp. 17-30. (*equal contribution) 
DOI: \url{https://doi.org/10.5334/tismir.39}\\\hline

\idIUI & \textbf{Kowald, D.*}, Lex, E.*, Schedl, M. (2020). Utilizing Human Memory Processes to Model Genre Preferences for Personalized Music Recommendations. In \textit{4th Workshop on Transparency and Explainability in Adaptive Systems through User Modeling Grounded in Psychological Theory (HUMANIZE @ ACM IUI'2020)}.  (*equal contribution) 
DOI: \url{https://doi.org/10.48550/arXiv.2003.10699}\\\hline

\idFNT & Lex, E., \textbf{Kowald, D.}, Seitlinger, P., Tran, T., Felfernig, A., Schedl, M. (2021). Psychology-informed Recommender Systems. \textit{Foundations and Trends in Information Retrieval}, 15:2, pp. 134–242. 
DOI: \url{https://doi.org/10.1561/1500000090}\\\hline

\idRecSysACTR & Moscati, M., Wallmann, C., Reiter-Haas, M., \textbf{Kowald, D.}, Lex, E., Schedl, M. (2023). Integrating the ACT-R Framework and Collaborative Filtering for Explainable Sequential Music Recommendation.  In \textit{Proceedings of the 17th ACM Conference on Recommender Systems (RecSys'2023)}, pp. 840–847. 
DOI: \url{https://doi.org/10.1145/3604915.3608838}\\\hline\hline

& \textbf{Privacy and Limited Preference Information in Recommender Systems}\\\hline

\idRecSysTRUST & Duricic, T., Lacic, E., \textbf{Kowald, D.}, Lex, E. (2018). Trust-Based Collaborative Filtering: Tackling the Cold Start Problem Using Regular Equivalence. In \textit{Proceedings of the 12th ACM Conference on Recommender Systems (RecSys'2018)}, pp. 446–450. 
DOI: \url{https://doi.org/10.1145/3240323.3240404}\\\hline

\idUMUAI & Lacic, E., Reiter-Haas, M., \textbf{Kowald, D.}, Dareddy, M., Cho, J., Lex, E. (2020). Using Autoencoders for Session-based Job Recommendations. \textit{User Modeling and User-Adapted Interaction}, 30, pp. 617–658. 
DOI: \url{https://doi.org/10.1007/s11257-020-09269-1}\\\hline

\idEcirMETA & Muellner, P., \textbf{Kowald, D.}, Lex, E. (2021). Robustness of Meta Matrix Factorization Against Strict Privacy Constraints. In \textit{Proceedings of the 43rd European Conference on Information Retrieval (ECIR'2021)}, pp. 107-119. 
DOI: \url{https://doi.org/10.1007/978-3-030-72240-1_8}\\\hline 

\idTIST & Muellner P., Lex, E., Schedl, M., \textbf{Kowald, D.} (2023). ReuseKNN: Neighborhood Reuse for Differentially-Private KNN-Based Recommendations. \textit{ACM Transactions on Intelligent Systems and Technology}, 14:5, pp. 1-29. 
DOI: \url{https://doi.org/10.1145/3608481}\\\hline 

\idFRONTPRI & Muellner P., Lex, E., Schedl, M., \textbf{Kowald, D.} (2023). Differential Privacy in Collaborative Filtering Recommender Systems: A Review.  \textit{Frontiers in Big Data}, 6:1249997, pp. 1-7. 
DOI: \url{https://doi.org/10.3389/fdata.2023.1249997}\\\\\\\hline 

& \textbf{Fairness and Popularity Bias in Recommender Systems}\\\hline

\idEcirPOP & \textbf{Kowald, D.}, Schedl, M., Lex, E. (2020). The Unfairness of Popularity Bias in Music Recommendation: A Reproducibility Study. In \textit{Proceedings of the 42nd European Conference on Information Retrieval (ECIR'2020)}, pp. 35-42. 
DOI: \url{https://doi.org/10.1007/978-3-030-45442-5_5}\\\hline 

\idEPJ & \textbf{Kowald, D.}, Muellner, P., Zangerle, E., Bauer, C., Schedl, M., Lex, E. (2021). Support the Underground: Characteristics of Beyond-Mainstream Music Listeners. \textit{EPJ Data Science}, 10:14. 
DOI: \url{https://doi.org/10.1140/epjds/s13688-021-00268-9}\\\hline

\idRecSysLBR & Lesota, O., Melchiorre, A., Rekabsaz, N., Brandl, S., \textbf{Kowald, D.}, Lex, E., Schedl, M. (2021). Analyzing Item Popularity Bias of Music Recommender Systems: Are Different Genders Equally Affected? In \textit{Proceedings of the 15th ACM Conference on Recommender Systems (RecSys'2021)}, pp. 601-606. 
DOI: \url{https://doi.org/10.1145/3460231.3478843}\\\hline 

\idBiasMEDIA & \textbf{Kowald, D.}, Lacic, E. (2022). Popularity Bias in Collaborative Filtering-Based Multimedia Recommender Systems. In \textit{Advances in Bias and Fairness in Information Retrieval (BIAS @ ECIR'2022)}. Communications in Computer and Information Science, vol. 1610, pp. 1-11. 
DOI: \url{https://doi.org/10.1007/978-3-031-09316-6_1}\\\hline 

\idEcirPRESSE & Lacic, E., Fadljevic, L., Weissenboeck, F., Lindstaedt, S., \textbf{Kowald, D.} (2022). What Drives Readership? An Online Study on User Interface Types and Popularity Bias Mitigation in News Article Recommendations. In \textit{Proceedings of the 44th European Conference on Information Retrieval (ECIR'2022)}, pp. 172-179. 
DOI: \url{https://doi.org/10.1007/978-3-030-99739-7_20}\\\hline 

\idBiasCALIBRATION & \textbf{Kowald, D.*}, Mayr, G.*, Schedl, M., Lex, E. (2023). A Study on Accuracy, Miscalibration, and Popularity Bias in Recommendations. In \textit{Advances in Bias and Fairness in Information Retrieval (BIAS @ ECIR'2023)}. Communications in Computer and Information Science, vol. 1840, pp. 1-16. (*equal contribution) 
DOI: \url{https://doi.org/10.1007/978-3-031-37249-0_1}\\\hline 

\idSCIREP & Scher, S., Kopeinik, S., Truegler, A., \textbf{Kowald, D.} (2023). Long-Term Dynamics of Fairness: Understanding the Impact of Data-Driven Targeted Help on Job Seekers. \textit{Nature Scientific Reports}, 13:1727.  
DOI: \url{https://doi.org/10.1038/s41598-023-28874-9}\\\hline 

\label{tab:papers}
\end{longtable}

I have contributed substantially to all 17 publications, and for 10 of these publications I am also either first or last author. The full texts of these publications can be found when following the respective DOI links in Table~\ref{tab:papers}. For a full list of my publications, please take a look at my Google Scholar profile:~\url{https://scholar.google.at/citations?user=qQ-L8rUAAAAJ&hl=en}.

\newpage


\chapter{Related Work and Background}
\label{c:relwork}

This chapter describes relevant research and background related to the scientific contributions of this habilitation. First, the main concepts of recommender systems are briefly outlined in Section~\ref{s:rel_main}, followed by relevant background with respect to transparency and cognitive models in recommender systems in Section~\ref{s:rel_transparency}. Next, the topic of privacy and limited preference information in recommender systems is briefly discussed in Section~\ref{s:rel_privacy_sparsity}. Finally, Section~\ref{s:rel_fairness_bias} gives a compact overview of fairness and popularity bias in recommender systems. This chapter also summarizes our own research related to these topics, which is then outlined in relation to the main publications of this habilitation in Chapter~\ref{c:contributions}.

\section{Main Concepts of Recommender Systems}
\label{s:rel_main}

This section gives a compact overview of recommender systems (i) algorithms, (ii) applications, and (iii) evaluation methods relevant to this habilitation.

\subsubsection{Recommender Systems Algorithms}

In general, there are three main categories of recommendation algorithms~\cite{adomavicius2005toward,ricci2010introduction}: (i) collaborative filtering (CF), (ii) content-based filtering (CBF), and (iii) hybrid approaches. This habilitation focuses on CF, but also investigates CBF. 

Typically, a user-based CF recommender system $\mathcal{R}^k$ generates an estimated rating score for a target user $u$ and a target item $i$ by utilizing the ratings $r_{n, i}$ of $k$ other users that have rated $i$, i.e., the $k$ nearest neighbors $N^k_{u, i}$~\cite{desrosiers2010comprehensive}. Therefore, this variant of CF is often referred to as \emph{UserKNN}, i.e., user-based $k$ nearest neighbors. Formally, the estimated rating score $\mathcal{R}^k(u, i)$ for $u$ and $i$ is given by:
\begin{equation}
   \mathcal{R}^k(u, i) = \frac{\sum_{n \in N^k_{u, i}} sim(u, n) \cdot r_{n, i}}{\sum_{n \in N^k_{u, i}} sim(u, n)}
   \label{eq:userknn}
\end{equation} 
where $sim(u, n)$ is the similarity between target user $u$ and neighbor $n$. 
For \emph{UserKNN}, the neighborhood $N^k_{u, i}$ used for generating recommendations for $u$ and $i$ comprises the $k$ most similar neighbors. More formally:
\begin{equation}
    N^k_{u, i} =\ \stackrel{k}{\argmax_{c \in U_i}} sim(u, c)
    \label{eq:userknn_neighborhood}
\end{equation}
where $U_i$ are all users that have rated $i$ and $sim$ is the similarity metric (e.g., Cosine or Pearson~\cite{benesty2009pearson}). There also exist variations of this algorithm suitable for item relevance prediction and for implicit user preferences (e.g., clicks)~\cite{jannach2018recommending}.

It is also possible to calculate similarities between items based on users' preferences of these items. This variant of CF is termed item-based CF (or \textit{ItemKNN}) and has advantages in cases when user profiles change quickly~\cite{sarwar2001item}. \textit{ItemKNN} was introduced as the main recommendation algorithm by Amazon.com~\cite{linden2003amazon} in 2003. For a comprehensive review of KNN-based CF methods, please see~\cite{nikolakopoulos2021trust}, and for a survey on CF with side information, please see~\cite{shi2014collaborative}. Another possibility to incorporate side and context information (e.g., time or location) is by utilizing context-aware recommender systems, as discussed in these works~\cite{adomavicius2005incorporating,adomavicius2010context,adomavicius2021context}.

Another variant of CF is \textit{matrix factorization} (MF), which follows the idea that a user's preferences can be efficiently represented in low-dimensional space~\cite{koren2009matrix,shi2010list}. The items are represented in the same low-dimensional space, which enables to generate recommendations by calculating the dot-product between the user and the item vectors. These vectors are often termed \textit{embeddings}, and can be calculated with techniques such as graph neural networks~\cite{sun2018recurrent}, recurrent neural networks~\cite{hidasi2018recurrent}, neural CF~\cite{he2017neural}, or autoencoders~\cite{zhang2020survey}. As described in Section~\ref{s:rel_privacy_sparsity}, in this habilitation, autoencoders are used to address the issue of limited preference information in session-based recommendations~\cite{jannach2017recurrent,li2023exploiting}. Furthermore, a neural CF approach~\cite{he2017neural} is used to study differentially-private recommendations. For a comparison of neural network and KNN-based methods, please see~\cite{ferrari2019we}.

The next type of algorithms, content-based filtering (CBF)~\cite{lops2010content}, utilizes content features of items (e.g., genres, title) to build item profiles to overcome the item cold-start problem (i.e., items with no user preference information). These item profiles are then matched with user profiles that also consist of content features of the consumed items~\cite{de2015semantics}. For representing content features, techniques such as LDA (Latent Dirichlet Allocation)~\cite{blei2003latent} can be used. 
CBF could suffer from a lack of novelty and diversity, since typically items are recommended that are similar to the items the user has consumed in the past. To overcome this issue, hybrid recommendation approaches~\cite{burke2002hybrid,burke2007hybrid,parra2014see,kouki2015hyper} combine CBF and CF to get the benefits of both worlds. There exist several ways to combine recommendation algorithms~\cite{jannach2010recommender}, including (i) monolithic, where collaborative and content information is combined in a single recommendation model, (ii) parallelized, where the results of different algorithms are combined using, e.g., a weighted approach, and (iii) pipelined, where one algorithm uses the results of another algorithm as input. 

\subsubsection{Recommender Systems Applications}

This habilitation focuses on four application areas for recommender systems, namely (i) tag recommendations, (ii) music recommendations, (iii) job recommendations, and (iv) news article recommendations. The following paragraphs briefly describe the particularities of these application areas. Tag recommendation systems aim to support users in finding descriptive tags (or keywords) for annotating Web resources~\cite{jaschke2007tag,marinho2008collaborative} (e.g., music tracks or tweets in Twitter). Previous research of this habilitation's author has shown that a user's choice of tags is affected by activation processes in human memory~\cite{kowald2015evaluating,kowald2016influence}, which can be utilized for a transparent design of tag recommendation models~\cite{kowald2017temporal} (see Section~\ref{s:rel_transparency}).

Similar to recommendations in other multimedia domains~\cite{deldjoo2020recommender} (e.g., movies or television items~\cite{masthoff2004group}), music recommender systems help users to navigate large content databases, and to find content that suit their taste~\cite{schedl2018current}. However, in contrast to movies or books, music has some distinguished properties that also affect the design of music recommendation algorithms~\cite{schedl2015music}: (i) music may be consumed repeatedly, while movies or books are typically consumed only once or a few times at maximum, (ii) music recommendations can be addressed on different abstraction levels including tracks, albums, artists and genres, (iii) rating data is relatively rare in the music domain, and thus implicit user preferences (e.g., listening events) are an important information source for recommender systems~\cite{dror2012yahoo}, and (iv) domain knowledge (e.g., musical sophistication) may have a high impact on how recommendations are perceived by the music listeners~\cite{jin2018effects}.   

Next, job recommender systems address a particular recommendation problem, in which open job positions should be matched with job candidates~\cite{abel2015we,abel2016recsys}. This differs to other recommendation application domains, since typically every open job position (i.e., the item) can be assigned to only one job candidate (i.e., the user), and vice versa~\cite{kenthapadi2017personalized}. Additionally, job portals (especially those that offer jobs to students and young talents) often provide the possibility to browse jobs anonymously~\cite{reiter2020heterogeneous,lacic2017beyond}, which then turns the job recommendation problem into a session-based recommendation problem~\cite{jannach2017recurrent,quadrana2017personalizing}. 
Limited preference information and anonymous user sessions are also issues of news article recommender systems~\cite{moniz2017framework,de2018news,hopfgartner2016benchmarking}. Via providing recommendations of currently relevant news articles that match session information (e.g., clicks) of the user, news portals aim to increase user engagement, and to turn anonymous readers into paying subscribers~\cite{abdollahpouri2021toward}. Finally, another particularity of news recommendations is the short lifetime of items, since many articles are only relevant for one day~\cite{ozgobek2014survey}. 

\subsubsection{Recommender Systems Evaluation Methods}

This habilitation considers both online and offline evaluation procedures of recommender systems. Both methods aim to compare the performance of two or more recommendation algorithms, but while online evaluation is performed in a live system, e.g., using A/B tests~\cite{garcin2014offline}, offline evaluation is performed using collected preferences, typically in the form of training and test sets~\cite{castells2022offline}. Another difference lies in the time when the user preference information is collected: whereas online evaluation collects user preferences after the recommendations are shown to the users, offline evaluation gathers user preferences (i.e., the ground truth data in the test set) before the recommendations are calculated~\cite{castells2023recommender,zangerle2022evaluating}. Online evaluation procedures then measure the actual performance using impact- or value-oriented measurements such as \textit{Click-Through-Rate (CTR)}~\cite{jannach2019measuring}. In contrast, offline evaluation procedures make use of relevance or performance metrics, which are often borrowed from the information retrieval research field~\cite{baeza1999modern,sanderson2010test}. 

With respect to offline evaluation metrics, this habitation investigates both accuracy and beyond-accuracy metrics. To measure  accuracy~\cite{cremonesi2010performance}, error-based metrics for rating prediction such as the \textit{Mean Absolute Error (MAE)}~\cite{willmott2005advantages}, and metrics for ranking quality such as \textit{Precision (P)}, \textit{Recall (R)}, \textit{F1-score (F1)}, \textit{Mean Reciprocal Rank (MRR)}, and \textit{Normalized Discounted Cumulative Gain (nDCG)}, have been proposed in the literature (e.g.,~\cite{herlocker2004evaluating}). 

After decades of accuracy-driven recommender systems evaluation procedures, the research community has argued that being accurate is not the only important objective for a recommender system, and has proposed a set of beyond-accuracy metrics~\cite{mcnee2006being,ge2010beyond,adomavicius2011improving,frontiers_gnn_2023}. Here, especially, the concepts of \textit{novelty} and \textit{diversity} are important~\cite{castells2021novelty}. Novelty describes the difference between the recommended items and a specific context, which could be the target user's item history or all users' item histories in the system~\cite{vargas2011rank}. The former, which is also referred to as personalized or user-based novelty, or unexpectedness~\cite{adamopoulos2014unexpectedness}, describes how different the recommendation list is from the items the target user has consumed in the past (i.e., the user item history). This concept is also related to serendipity, which, in addition, takes the relevance of the recommended items into account~\cite{chen2019serendipity}. The latter, which is also referred to as long-tail novelty or system-based novelty~\cite{vargas2011rank}, measures the rarity or inverse popularity of the recommended items~\cite{celma2008new}. This concept is also related to evaluating fairness and popularity bias of recommendations, which is described in more detail in Section~\ref{s:rel_fairness_bias}. 

All methods and metrics discussed so far solely evaluate the recommender system from a user's, or consumer's, perspective. However, in recent years, the multi-stakeholder nature of recommender systems has been highlighted, which not only takes the users, but also the item providers (and maybe other stakeholders like the system operators) into account~\cite{abdollahpouri2017recommender,abdollahpouri2020multistakeholder,smith2023many,smith2024recommend}. Here, especially integrating and evaluating item provider constraints is becoming an important research topic~\cite{surer2018multistakeholder}, and is also related to multi-sided fairness aspects of recommender systems~\cite{burke2018balanced,sonboli2022multisided}. Finally, the reproducibility of recommender systems evaluation procedures is another important and timely topic~\cite{cremonesi2021progress}. Here, the adequate documentation and sharing of source-code and dataset samples used in the evaluation process is a key aspect of reproducibility~\cite{beel2016towards}. Please see Section~\ref{sec:cont_summary}, for a discussion of reproducibility aspects related to the contributions of this habilitation. 

\findingbox{This habilitation studies a wide range of recommendation algorithms and applications such as tag, music, job, and news article recommendations. Additionally, we investigate both accuracy and beyond-accuracy metrics, and both offline and online evaluation settings. Finally, we discuss reproducibility aspects of the scientific contributions of this habilitation, and provide code and data resources to foster reproducibility.}

\section{Transparency and Cognitive Models in Recommender Systems}
\label{s:rel_transparency}

This habilitation investigates transparency aspects of recommender systems by following principles of psychology and human cognition for a transparent design process of recommendation algorithms. Another possibility to enhance transparency in recommender systems is by providing explanations for recommendations, which is not investigated in this habilitation. For the field of explainability in recommender systems, please see~\cite{tintarev2007survey,tintarev2010designing,tintarev2015explaining,nunes2017systematic}.

\subsection{The Role of Psychology in Recommender Systems}
\label{s:rel_psych}

Already, early research in the field of recommender systems was influenced by the fact that humans' decision-making processes are impacted by their social surroundings, which also motivated the implementation of the first collaborative filtering-based recommendation algorithms~\cite{resnick1994grouplens,resnick1997recommender}. In order to create more human-centric recommendations, additional psychological characteristics of users were incorporated in the design and implementation process of recommender systems~\cite{tran2021humanized,atas2021towards}. For example, insights from decision psychology~\cite{lee2002effects} were used to study serial position and anchoring effects in recommendations~\cite{felfernig2007persuasive,adomavicius2013recommender,stettinger2015anchoring}, and to show that users are more likely to remember items at the beginning (i.e, \textit{primacy} effect) and the end (i.e., \textit{recency} effect) of a list~\cite{stettinger2015counteracting}. Related research also investigated how to incorporate aspects such as personality~\cite{tkalcic2015personality}, and affect, e.g., emotion~\cite{gonzalez2002emotional} or satisfaction~\cite{masthoff2005pursuit,masthoff2006pursuit}, into the recommendation process.

Based on these lines of research, we survey and categorize related work at the intersection of psychology and recommender systems. We term this type of recommender systems \textit{psychology-informed recommender systems}~\cite{lex2021psychology}, and we identify three main areas: (i) cognitive (or cognition)-inspired, (ii) personality-aware, and (iii) affect-aware recommender systems. Additionally, we connect these areas to aspects of human decision-making, and to aspects of human-centric evaluation design of recommender systems. This habilitation focuses on the first area, namely cognitive-inspired recommendations based on human memory theory, which is described in more detail in the following section.

\findingbox{We highlight the usefulness of incorporating the underlying psychological constructs and theories into a transparent design process of recommender systems. We term this type of recommender system \textit{psychology-informed recommender system}, and categorize it into three types.}

\subsection{Cognitive-inspired Recommendations}
\label{s:rel_cognitive}

This habilitation investigates two cognitive-inspired recommendation approaches: one based on human episodic memory, and another one based on activation processes in human memory. Other types of cognition-aware recommendation approaches, such as stereotype-based recommendations~\cite{rich1979user}, categorization-based recommendations~\cite{seitlinger2013recommending}, or attention-based user models~\cite{seitlinger2015attention}, are discussed in~\cite{lex2021psychology}. 

\subsubsection{Recommendations based on Human Episodic Memory}

Human episodic memory is the memory of personally experienced events that occurred in a specific context (e.g., a particular day or place, or a given categorization)~\cite{tulving1993episodic}. The contextual information is essential for retrieving these events. MINERVA2~\cite{hintzman1984minerva} is a model that accounts for episodic memory-based human behavior such as categorization~\cite{hintzman1986schema}, and recognition~\cite{hintzman1988judgments}. MINERVA2 distinguishes between a long-term or secondary memory that holds the episodic memory traces (i.e., the events along with the context information), and a working or primary memory that communicates with the secondary memory by sending retrieval cues (e.g., current context information), and receiving matching events.

In our own research~\cite{seitlinger2018balancing,kowald2018impact}, we employ MINERVA2 to implement a tag recommendation algorithm called \textit{Search of Memory (SoMe)}. \textit{SoMe} mimics a user's search of memory when assigning tags to bookmark a Web resource. Therefore, we encode episodic memory traces using the categories assigned to previously bookmarked Web resources of this user. Specifically, \textit{SoMe} implements MINERVA2's distinction between the primary and secondary memory in a way that the primary memory represents the Web resource to be tagged in terms of the resource's categories, and to search the secondary memory for tags that are assigned to Web resources with similar categories. These tags are then recommended to the user. Via user studies, we find that \textit{SoMe} provides higher tag recommendation acceptance than a popularity-based baseline approach~\cite{seitlinger2018balancing,kowald2018impact} (see Section~\ref{sec:cont_transparency}).

\subsubsection{Recommendations based on Activation Processes in Human Memory}

Human memory is very efficient in making memory units quickly available when they are needed~\cite{bettman1980effects,park1981familiarity}. More formally, human memory tunes the activation of its units to statistical regularities of the current context and environment~\cite{anderson1991reflections}. These so-called activation processes in human memory are formalized in the cognitive architecture ACT-R~\cite{anderson1997act}. ACT-R is short for ``Adaptive Control of Thought – Rational'', and differs between two long-term memory modules: (i) declarative memory, which holds factual knowledge (i.e., what something is), and (ii) procedural memory, which consists of action sequences (i.e., how to do something)~\cite{anderson2004integrated}. 

This habilitation focuses on the declarative memory module, which contains the \textit{activation equation} of human memory. The \textit{activation equation} determines the usefulness, i.e., the activation level $A_i$, of a memory unit $i$ (e.g., a specific item or item category the user has interacted with in the past) for a user $u$ in the current context. It combines a \textit{base-level} activation with an \textit{associative} activation, which depends on the weight $W_j$, and the strength of association $S_{j,i}$~\cite{anderson2004integrated}:
\begin{equation} \label{eq:A}
    A_i~=~B_i + \sum_j{W_j \cdot S_{j,i}}
\end{equation}
where $B_i$ represents the \textit{base-level} activation of $i$, which quantifies its general usefulness by considering how frequently and recently it has been used in the past. It is defined by the \textit{base-level learning (BLL) equation}~\cite{anderson1991reflections}: 
\begin{equation} \label{eq:bll}
    B_i~=~ln\left(\sum\limits_{j~=~1}\limits^{n}{t_{j}^{-d}}\right)
\end{equation}
where $n$ is the frequency of $i$'s occurrences in the past (i.e., how often $u$ has interacted with $i$), and $t_j$ is the time since the $j^{th}$ occurrence of $i$ (i.e., the recency of $i$). The exponent $d$ accounts for the time-dependent decay of item exposure, which means that each unit's activation level decreases in time according to a power function. The second part of Equation~\ref{eq:A} represents the \emph{associative activation} that tunes $B_i$ to the current context. The current context can be defined by any contextual element $j$ that is relevant to the current situation, and  via learned associations, the contextual elements can increase $i$'s activation.

\begin{figure}[t!] 
    \centering
    \includegraphics[width=0.88\textwidth]{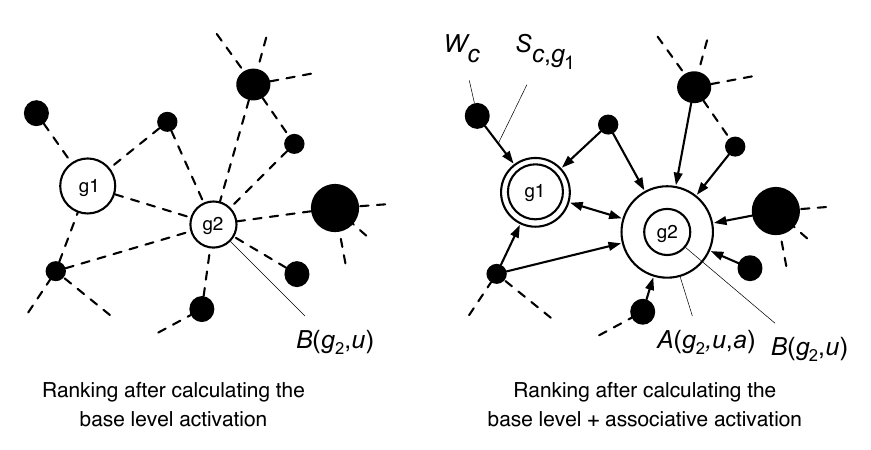}
    \caption{An example illustrating the difference between the \textit{BLL equation} (left panel) and the \textit{activation equation} (right panel). Here, unfilled nodes represent target genres $g_1$ and $g_2$, and black nodes represent contextual genres. For $g_1$ and $g_2$, the node sizes represent the activation levels, and for the contextual genres, the node sizes represent the weights $W_c$. The association strength $S_{c,g}$ is represented by each edge's length. We see a different ranking of the genres in the two settings, which illustrates the importance of the associative activation~\cite{kowald2015refining,trattner2016modeling,kowald2020utilizing}.}
    \label{fig:example}
\end{figure}

Figure~\ref{fig:example} illustrates the difference between the base-level activation and the associative activation in the case of a music recommendation system that aims to rank relevant music genres for a given user. The left panel shows the ranking of two genres $g_1$ and $g_2$ according to the \textit{BLL equation}. Here, $g_1$ would have a higher activation level than $g_2$ based on past usage frequency and recency. The right panel shows the ranking of these genres according to the \textit{activation equation}, which also takes associations with contextual genres into account (e.g., music genres that are relevant in the current situation). Using the combined base-level and associate activation, the ranking changes, and $g_2$ would have a higher activation level than $g_1$~\cite{kowald2020utilizing}. 
The declarative memory module also contains some additional components. One example is the valuation component~\cite{juvina2018modeling}, which determines the \textit{value} attributed by $u$ to $i$ (e.g., interaction time or frequency~\cite{reiter2021predicting}).

In our research, we use the \textit{BLL equation} and \textit{activation equation} for a transparent design, implementation, and evaluation process of two music recommendation algorithms~\cite{lex2020modeling,kowald2020utilizing}. We show that these cognitive-inspired approaches outperform related baselines in terms of recommendation accuracy. Additionally, we illustrate to what extent the components of ACT-R contribute to the generation of the music recommendation lists~\cite{recsys_actr_2023}. In a recently accepted paper~\cite{humanizebook}, we discuss transparency aspects of additional components of ACT-R.

\findingbox{We use models of human episodic memory (i.e., MINERVA2), and activation process in human memory (i.e., ACT-R) for a transparent design, implementation, and evaluation process of recommendation algorithms. We also illustrate to what extent the components of ACT-R (e.g., \textit{BLL}) contribute to the generation of the recommendation lists.}

\section{Privacy and Limited Preference Information in Recommender Systems}
\label{s:rel_privacy_sparsity}

This section gives an overview of privacy-aware recommendations. Since the users' privacy concerns could also lead to the limited availability of preference information (e.g., users disclose their preferences, or do not sign in to the system), this section also gives an overview of session-based and cold-start recommendations.

\subsection{Privacy-aware Recommendations}
\label{s:rel_privacy}

In terms of privacy, this habilitation focuses on differentially-private recommendations. This section also briefly discusses privacy aspects of recommender systems.

\subsubsection{Privacy Aspects of Recommender Systems}

Recommender systems need to store and process user preference information, which could lead to potential privacy risks to its users~\cite{friedman2015privacy}. This includes the inference of private information. Here, related research has shown that inference attacks can be used to derive a user's sensitive information (e.g., gender~\cite{slokom2021towards}) based on the information shared with the recommender system~\cite{weinsberg2012blurme,jeckmans2013privacy,beigi2020survey}. For example, in $k$-nearest neighbor-based recommender systems, the use of neighbors' preference information in the recommendation process can pose a privacy risk to the neighbors~\cite{ramakrishnan2001being,zhang2021membership}. In this way, the preference information of the neighbors can be uncovered, or the neighbors' identities (or sensitive attributes) can be revealed. Other inference attacks in recommender systems work by generating fake users, i.e., sybils, based on the limited knowledge of a victim's preferences.
These sybils isolate the victim utilized as a neighbor, and compromise its privacy~\cite{calandrino2011you}.

Different privacy-preserving technologies have been used to mitigate the users' privacy risks, including \textit{homomorphic encryption}, \textit{federated learning}, and \textit{differential privacy}. While \textit{homomorphic encryption} techniques aim to generate privacy-aware recommendations by employing encrypted user preference information~\cite{zhang2021privacy}, \textit{federated learning} techniques build on the assumption that sensitive user information should never leave the user's device~\cite{yang2019federated,anelli2021federank}. Finally, \textit{differential privacy} protects the users by introducing noise into the recommendation process~\cite{dwork2008differential}.

Our own research focuses on using \textit{differential privacy}. 
Additionally, we study how limiting the preference information of users can help to increase privacy. Therefore, we use the concept of \textit{meta learning}~\cite{lin2020meta} to calculate recommendations based on a minimal amount of user preference information. With this, we study privacy constraints of users (e.g., willingness to share preference information)~\cite{muellner2021robustness}. We find that users with small profiles can afford a higher degree of privacy than users with large profiles, and that \textit{meta learning} is helpful for increasing the robustness against the users' privacy constraints (see Section~\ref{sec:cont_privacy}).

\subsubsection{Differentially-Private Recommendations}

The aim of differentially-private recommendations is to inject randomness and noise into the recommendation calculation process to mitigate the inference risk of users' preference information~\cite{mcsherry2009differentially,friedman2016differential}. This habilitation focuses on a specific attack, which can be addressed by using \textit{differential privacy}. Here, a user with malicious intent, i.e., the \textit{adversary} $a$, tries to infer preference information (here, rating scores) of a specific neighbor $n$ in user-based $k$-nearest neighbor CF (i.e., \textit{UserKNN})~\cite{calandrino2011you}. In this attack scenario, the adversary $a$ has some prior knowledge about $n$, such as publicly available rating information $P$ of $n$ that could have been inferred from, e.g., product reviews. Using $P$, $a$ modifies its own user profile $R_a$ such that it (partially) matches $n$'s profile, which increases the likelihood of $n$ being used as a neighbor for calculating $a$'s recommendations. With this, $a$ queries estimated rating scores from the recommender system, i.e., $\mathcal{R}^k(a) = \{\mathcal{R}^k(a, i_1), \mathcal{R}^k(a, i_2), \dots, \mathcal{R}^k(a, i_l)\}$, where $\mathcal{R}^k(a, i_j)$ is the estimated rating score for item $i_j \in Q_a$, and $Q_a$ is the set of $a$'s rating queries. Then $a$ aims to infer rating information $r_{n, i_j}$ of a neighbor $n$ for item $i_j$ used to generate the estimated rating scores. More formally, this is given by:
\begin{equation}
    \label{eq:attack}
    Pr[r_{n, i_1}, r_{n, i_2}, \dots, r_{n, i_l} | \mathcal{R}^k(a, i_1), \mathcal{R}^k(a, i_2), \dots, \mathcal{R}^k(a, i_l), P \cup R_a]
\end{equation}

To mitigate the inference risk of $n$'s rating information, different variants of \textit{differential privacy} such as the \textit{Laplace input perturbation}~\cite{dwork2014algorithmic} or \textit{plausible deniability}~\cite{bindschaedler2017plausible} can be used. This habilitation utilizes \textit{randomized responses}~\cite{warner1965randomized} to establish \textit{plausible deniability}. Specifically, a privacy mechanism $m_{DP}$ is applied to the neighbors' ratings to generate the differentially-private set of ratings $\Tilde{R}$:
\begin{equation}
    \Tilde{R} = \{m_{DP}(r_{n, i}): n \in N^k_{u, i}\} 
\end{equation}
Via \textit{randomized responses}, neighbors can plausibly deny that their real rating was used in the recommendation process.
In detail, the privacy mechanism $m_{DP}$ flips a fair coin, and if the coin is heads, the neighbor's real rating is used in the recommendation calculation. If the coin is tails, $m_{DP}$ flips a second fair coin to decide whether the neighbor's real rating, or a random rating drawn from a uniform distribution over the range of ratings, is used.
With this, the adversary $a$ does not know if the utilized rating is real, or random, which leads to the guarantees of \textit{differential privacy}~\cite{dwork2014algorithmic}. However, the randomness introduced to the users' preference information typically leads to accuracy drops, and thus also to a fundamental trade-off between accuracy and privacy~\cite{berkovsky2012impact}. 

\begin{figure}[t!]
    \centering
    \includegraphics[width=0.95\linewidth]{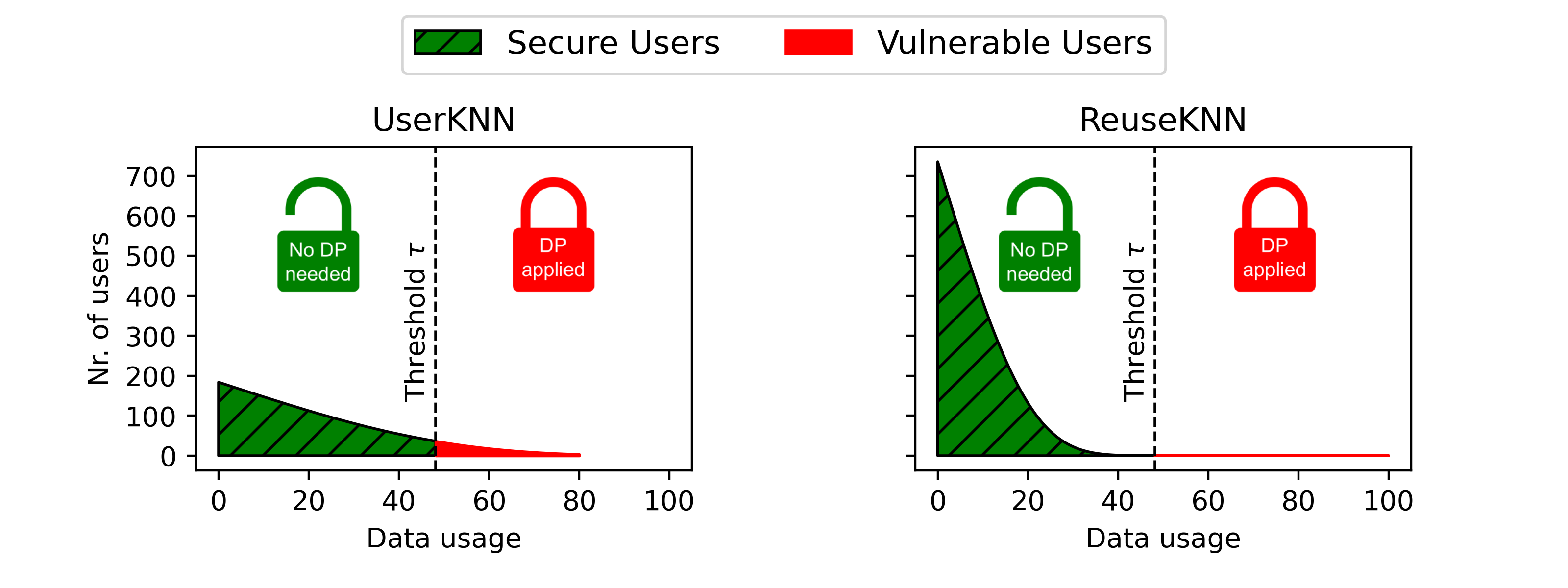}
    \caption{Schematic illustration of the data usage (i.e., how often a user is used as a neighbor) distribution of \emph{UserKNN} and \emph{ReuseKNN}. 
    \emph{ReuseKNN} increases the number of \textit{secure} users (green, no \textit{differential privacy} needed) and decreases the number of \textit{vulnerable} users (red, \textit{differential privacy} needs to be applied) compared to \emph{UserKNN}. The dashed line illustrates the data usage threshold $\tau$, a parameter to adjust the maximum data usage for users to be treated as secure.\vspace{-2mm}}
    \label{fig:reuseknn_approach}
\end{figure}

In our research, we address this accuracy-privacy trade-off by proposing a novel differentially-private recommendation approach termed \textit{ReuseKNN}~\cite{tist_dp_2023}. \textit{ReuseKNN} aims to reduce the number of users that need to be protected via \textit{differential privacy} by employing an efficient neighborhood reuse concept. With this, the majority of users (we call them \textit{secure} users) are rarely used in the recommendation process and thus, do not need protection, while some highly reusable users (we call them \textit{vulnerable} users) can be protected with \textit{differential privacy}. Figure~\ref{fig:reuseknn_approach} schematically illustrates our approach, and shows that the fraction of \textit{secure} users is substantially larger in the case of \textit{ReuseKNN} compared to traditional \textit{UserKNN}. We also find that this leads to higher recommendation accuracy compared to a fully differentially-private recommender system (see Section~\ref{sec:cont_privacy}). Additionally, we survey, analyze, and categorize the use of \textit{differential privacy} in 26 papers published in recommender systems-relevant venues~\cite{frontiers_privacy_2023}.

\findingbox{We study varying privacy constraints of users, e.g., the willingness to share preferences with the recommender system. Additionally, we address the privacy-accuracy trade-off in differentially-private recommendations by employing a neighborhood reuse concept, and survey and categorize the literature on using \textit{differential privacy} in collaborative filtering.}
\vspace{-2mm}
\subsection{Limited Availability of User Preference Information}
\label{s:rel_sparse_cold}

Data protection initiatives as well as the users' privacy concerns in recommender systems can lead to the limited availability of preference information~\cite{knijnenburg2013making,larson2017towards,valdez2019users,biega2020operationalizing,mehdy2021privacy}. This habilitation investigates this issue in session-based and cold-start recommendation settings, which are discussed in this section.

\subsubsection{Session-based Recommendations}

Session-based recommender system aim to provide meaningful recommendations in cases where long-term user preferences, or user histories, are not available (e.g., due to users' privacy concerns, or when users do not sign in to the system). The input of a session-based recommender system consists of a typically short item sequence that is observed in the current user session~\cite{ludewig2018evaluation,jannach2022session}. Different algorithms for session-based recommendations have been proposed, including methods based on $k$-nearest neighbors~\cite{jannach2017recurrent} or recurrent neural networks~\cite{hidasi2018recurrent}. Session-based recommender systems are related to sequence-aware and sequential recommender systems~\cite{kang2018self}, which are not covered in this habilitation. Please see~\cite{quadrana2018sequence} for a detailed overview of sequence-aware and sequential recommendations. 

In our research, we employ autoencoders, a specific type of neural network for reducing the dimensionality of data~\cite{kramer1991nonlinear}, to infer latent session representations, and to generate session-based job recommendations. Specifically, we find that variational autoencoders provide the best results across a set of accuracy and beyond-accuracy evaluation metrics (e.g., system and session-based novelty)~\cite{lacic2020using}.

\subsubsection{The User Cold-Start Problem}

The user cold-start problem in recommender systems refers to users that have interacted with only a few or even with no items at all, i.e., users with limited availability of preference information~\cite{schein2002methods}. Related research has proposed different methods to address the user cold-start problem, including simple popularity-based and unpersonalized approaches~\cite{schein2002methods}, location-aware recommendations~\cite{lacic2015tackling}, and trust-based recommendations~\cite{massa2007trust,eirinaki2013trust}. This habilitation focuses on trust-based recommendations, which exploit trust statements between users (e.g., user $A$ trusts user $B$) to create trust networks, and to calculate CF-based recommendations using the connections in these trust networks~\cite{o2005trust,guo2015trustsvd,guo2016novel}.

In our research, we employ network measures such as \textit{regular equivalence}~\cite{helic2014regular} to calculate trust-based recommendations for cold-start users. Via \textit{regular equivalence}, we do not only find neighbors that share the same trust connections, but also neighbors that have similar structural roles in the trust network (e.g., users that are only connected to influential nodes in the network). We find that our approach outperforms related methods based on, e.g., Jaccard similarity~\cite{duricic2018trust}.

\findingbox{We address the issue of limited availability of user preference information (e.g., due to users' privacy constraints) in session-based and cold-start recommendation settings. We demonstrate the usefulness of variational autoencoders for session-based job recommender systems. Furthermore, we address the user cold-start problem by employing trust-based recommendations using network measures such as \textit{regular equivalence}.}

\section{Fairness and Popularity Bias in Recommender System}
\label{s:rel_fairness_bias}

This section gives a brief overview of fairness in algorithmic decision support, and outlines research on popularity bias in recommender systems. For more detailed reviews on fairness-aware recommender systems, please see~\cite{yang2017measuring,ekstrand2022fairness,deldjoo2023fairness,wang2023survey}.

\subsection{Fairness in Algorithmic Decision Support}
\label{s:rel_fairness}

Fairness in algorithmic decision support and in machine learning applications has gained a lot of attention in recent years, and has been studied especially for binary classification problems~\cite{kusner2017counterfactual,berk2018fairness,mehrabi2021survey}. In this problem setting, $Y$ denotes the real outcome to be predicted by the classifier (e.g., the class label, for example if a job applicant has been put into a high- or low-prospect group), and $A$ is the set of protected attributes of an individual, thus the attributes that one must not discriminate against (e.g., gender or race). Furthermore, $X$ denotes non-protected attributes of an individual, and $\hat{Y}$ is the predictor of $Y$ (e.g., to predict to which class the individual belongs), which could depend on $X$ and $A$. Different definitions of fairness were proposed for such a setting in the literature. 

For example, \textit{fairness through unawareness} is satisfied if the predictor $\hat{Y}$ only depends on $X$ and not on $A$ to predict $Y$, i.e., $\hat{Y}: X \rightarrow Y$. Although this fairness definition seems to be compelling and simple to implement, it was shown that it is not sufficient in the area of algorithmic decision support since elements of $X$ may contain hidden discriminatory information of $A$ (e.g., race may correlate with the place of residence)~\cite{hajian2012methodology}. Another definition is based on \textit{individual fairness}~\cite{dwork2012fairness}. Given that we have a distance metric $d(i, j)$, if two individuals $i$ (with $X_i$ and $A_i$) and $j$ (with $X_j$ and $A_j$) are similar according to this metric (so $d(i, j)$ is small), then also their predicted outcomes should be similar: $\hat{Y}(X_i, A_i) \approx \hat{Y}(X_y, A_y)$. One drawback of \textit{individual fairness} is that the definition of $d(i, j)$ requires detailed information of the individuals as well as detailed domain knowledge.

Apart from that, the literature has also provided different definitions for \textit{group fairness}. According to the \textit{statistical parity} (or demographic parity) definition~\cite{barocas2016big}, fairness is given if the positive outcome proportion of the predictor $P(\hat{Y} = 1)$ is equal for all $A$, which, in the binary case with $A \in \{0, 1\}$, is given by:
\begin{equation}
    P(\hat{Y} = 1 | A = 0) = P(\hat{Y} = 1 | A = 1)
\end{equation}
Legally, this metric is often related to the 4/5th rule~\cite{greenberg1979analysis}. This means that the positive outcome ratio between the protected group (i.e., $A = 0$) and the privileged group (i.e., $A = 1$) should be at least 0.8. For example, if the privileged group has a positive outcome proportion of 50\%, then the protected group should have a positive outcome proportion of at least 40\%. The downside of this metric is that it does not depend on the real outcome $Y$ (only on the predictions $\hat{Y}$).

In contrast, \textit{equality of opportunity}~\cite{hardt2016equality} also takes the real outcome $Y$ into account. The idea is that individuals of the privileged and individuals of the protected group should have equal chance of getting a positive outcome, assuming that the individuals of the groups are qualified for this positive outcome. This can be measured via the true positive rate, which is given by:
\begin{equation}
    P(\hat{Y} = 1 | Y = 1, A = 0) = P(\hat{Y} = 1 | Y = 1, A = 1)
\end{equation}
\textit{Equality of opportunity} can also be defined using the false negative rate~\cite{scher2023modelling}. Additionally, \textit{equalized odds} is a stricter variant of \textit{equality of opportunity} that requires that both the true positive rate and the false positive rate are equal~\cite{verma2018fairness}. Research has also found a trade-off between \textit{individual} and \textit{group fairness}~\cite{binns2020apparent}.

In our research, we employ some of these definitions and adjust them to study long-term dynamics of fairness in algorithmic decision support. Therefore, we develop an agent-based model and evaluate it in a labor market setting~\cite{scher2023modelling}. We find that there is a trade-off between different long-term fairness goals, which validates the aforementioned \textit{individual} and \textit{group fairness} trade-off (see Section~\ref{sec:cont_fairness}). 
Although, this work does not directly study recommender systems, it sheds light on the usefulness of agent-based modeling for studying algorithmic fairness in the long-term, which is also relevant for the research field of recommender systems.

\findingbox{We study long-term fairness dynamics in algorithmic decision support in a labor market setting using agent-based modeling techniques. We highlight the trade-off between different long-term fairness goals in such a setting (i.e., \textit{individual} and \textit{group fairness}).}

\subsection{Measuring, Understanding, and Mitigating Popularity Bias}
\label{s:rel_popbias}

In this section, metrics to measure and understand popularity bias, and methods to mitigate popularity bias in recommender systems are briefly discussed.

\subsubsection{Popularity Bias Metrics}

Research has shown that recommendation algorithms (especially those based on CF) are biased towards popularity, which leads to the overrepresentation of popular items in the recommendation lists~\cite{ekstrand2018all,elahi2021investigating}. This also leads to the underrepresentation of unpopular items (long-tail items) in the recommendation lists~\cite{brynjolfsson2006niches,park2008long}. The literature has proposed different metrics to measure and understand popularity bias from the item and user perspective~\cite{ahanger2022popularity,klimashevskaia2023survey}. This habilitation focuses on three specific ways to measure inconsistencies between user groups with respect to popularity bias: (i) accuracy differences between user groups, (ii) \textit{miscalibration}, and (iii) \textit{popularity lift}. While the first one simply requires comparing the average recommendation accuracy between the groups, \textit{miscalibration} and \textit{popularity lift} are more complex to calculate. Additionally, via \textit{skewness} and \textit{kurtosis}, we measure the asymmetry and ``tailedness'' of the popularity distributions~\cite{bellogin2017statistical}. 

In general, \textit{calibration} quantifies the similarity of a genre spectrum between a user profile $p$ and a list of recommendations $q$~\cite{steck2018}. For example, if a user has consumed 80\% of rock music and 20\% of pop music in the past, then a \textit{calibrated} recommendation list should also contain this genre distribution. Although this metric is not a popularity bias metric by definition, it is often used to measure and understand popularity bias in recommendations~\cite{abdollahpouri2019impact,abdollahpouri2020connection}. The definition of \textit{calibration} can be reinterpreted in the form of \textit{miscalibration}, i.e., the deviation between $p$ and $q$~\cite{lin2020}. This deviation is calculated using the \textit{Kullback-Leibler (KL)} divergence between the distribution of genres in $p$, i.e., $p(c|u)$, and the distribution of genres in $q$, i.e., $q(c|u)$. More formally, for user $u$, this is given by:
\begin{align}
KL(p||q) = \sum_{c \in C} p(c|u) \log \frac{p(c|u)}{q(c|u)}
\end{align}
Here, $C$ is the set of all genres in a given dataset. Therefore, $KL(p||q) = 0$ means perfect \textit{calibration}, and higher $KL(p||q)$ values (i.e., close to 1) mean \textit{miscalibrated} recommendations. The $KL(p||q)$ values can be averaged for a given group $g$. 

In contrast, \textit{popularity lift} measures to what extent recommendation algorithms amplify the popularity bias inherent in the user profiles~\cite{abdollahpouri2019unfairness,abdollahpouri2020connection}. Thus, this metric quantifies the disproportionate recommendation of popular items for a given user group $g$. \textit{Popularity lift} is based on the group average popularity $GAP_p(g)$, which is defined as the average popularity of the items in the user profiles $p$ of group $g$. Similarly, $GAP_q(g)$ is the average popularity of the recommended items for all users of the group $g$. Taken together:
\begin{align}
PL(g) = \frac{GAP_q(g) - GAP_p(g)}{GAP_p(g)}
\end{align}
$PL(g) > 0$ means that $g$'s recommendations are too popular, $PL(g) < 0$ means that $g$'s recommendations are too unpopular, and $PL(g) = 0$ is the ideal value. 

In our research, we use these metrics to study popularity bias in recommender systems~\cite{kowald2020unfairness,kowald2022popularity,ecir_bias_2023}. We find that ``niche'' users interested in unpopular content receive worse recommendation quality than users interested in popular content. We study the characteristics of these ``niche'' users in the field of music recommendations, and identify subgroups that also differ in the recommendation quality they receive~\cite{kowald2021support}. Finally, we also find that music recommendation algorithms could intensify the popularity bias for the group of female users~\cite{lesota2021analyzing}.

\subsubsection{Popularity Bias Mitigation}

Research has proposed different methods to mitigate bias in algorithms, including pre-, in-, and post-processing methods~\cite{ntoutsi2020bias}. In the field of recommender systems, especially in-processing and post-processing techniques are used to mitigate popularity bias. Here, in-processing algorithms aim to adjust the recommendation calculation procedure, and to correct the popularity bias using, e.g., \textit{calibration}-based techniques~\cite{abdollahpouri2021user,klimashevskaia2022mitigating}. In contrast, post-processing techniques do not change the recommendation algorithm itself, but the generated recommendation list by using, e.g., re-ranking techniques~\cite{antikacioglu2017post,abdollahpouri2019managing}. Typically, in-processing techniques are the most complex ones to implement, since the underlying algorithm needs to be adapted. However, they are efficient with respect to computational costs. In contrast, one drawback of post-processing techniques is the computational inefficiency of these methods due to the high computational complexity of item re-ranking. However, they can be applied to any given item ranking independent of the underlying algorithm~\cite{chen2023bias}. Finally, the use of content-based recommendation algorithms~\cite{lops2010content,de2015semantics} is another possibility to address popularity bias in recommender systems due to their independence of user preference information~\cite{abdollahpouri2017controlling,conext_dataeco_2022}. 

In our research, we study popularity bias mitigation in news article recommender systems for both subscribed users and anonymous session users utilizing content-based recommendations~\cite{lacic2022drives}. In an online study that we have conducted together with the Austrian news platform \textit{DiePresse}, we find that personalized and content-based recommendations lead to a more balanced news article readership distribution compared to purely popularity-based recommendations. Thus, we find that readers are not only interested in the most popular and recent news articles, but also in long-tail articles if they match the user preference history, or the preferences tracked in the current session (see Section~\ref{sec:cont_fairness}).

\findingbox{We analyze popularity bias in collaborative filtering-based recommender systems, and find that ``niche'' users interested in unpopular content receive worse recommendation accuracy than users interested in popular content. Thus, this ``niche'' user group is treated in an unfair way by collaborative filtering-based recommender systems. Furthermore, we analyze the characteristics of these users, and study popularity bias mitigation in news article recommender systems using content-based recommendations.}

Please note that the aim of this ``Related Work and Background'' chapter has not been to give a comprehensive review of the various research fields mentioned, but rather to discuss the research and background related to the scientific contributions and publications of this habilitation described in the next chapter.

\newpage


\chapter{Scientific Contributions}
\label{c:contributions}

This chapter describes the scientific contributions of this habilitation according to the three research topics that are investigated: (i) transparency and cognitive models (Section~\ref{sec:cont_transparency}), (ii) privacy and limited preference information (Section~\ref{sec:cont_privacy}), and (iii) fairness and popularity bias (Section~\ref{sec:cont_fairness}) in recommender systems. Therefore, the 17 publications listed in Table~\ref{tab:papers} are categorized into 7 scientific contributions. For research topics (i) and (ii), this leads to two contributions each, and for research topic (iii), this leads to three contributions, since this topic also covers the most publications of this habilitation.  
Additionally, Section~\ref{sec:cont_summary} summarizes the scientific contributions, and elaborates on reproducibility aspects.

\section{Transparency and Cognitive Models in Recommender Systems}
\label{sec:cont_transparency}

This section summarizes our research on transparency aspects of recommendations by using cognitive modeling techniques. It contains three studies employing cognitive models for a transparent design process of tag and music recommendation algorithms~\idHCI~\idTISMIR~\idIUI, and one survey and categorization of psychology-informed recommender systems~\idFNT~(\textit{Contribution 1}). Additionally, one study illustrates to what extent the components of the cognitive model ACT-R contribute to the generation of music recommendation lists~\idRecSysACTR~(\textit{Contribution 2}).

\subsubsection{Contribution 1: Using Cognitive Models for a Transparent Design and Implementation Process of Recommender Systems (2018-2021)}
\vspace{-1mm}
\idHCI~introduces a tag recommendation algorithm termed \textit{SoMe (Search of Memory)} based on \textit{MINERVA2}~\cite{hintzman1984minerva}, which is a model of human episodic memory (see Section \ref{s:rel_cognitive}). We implement \textit{SoMe} using our \textit{TagRec} framework~\cite{kowald2014tagrec,kowald2017tagrec}, and evaluate it in an online study with 18 participants. During the four-weeks study, the participants had to investigate a specific topic (i.e., ``designing workplaces that inspire people'') by collecting and tagging three topic-related Web resources per week. For this, the participants were supported with a social bookmarking user interface (based on the \textit{KnowBrain} tool~\cite{dennerlein2015knowbrain}) that contained support via tag recommendations. Here, the participants randomly received tag recommendations calculated via \textit{SoMe} or via a conventional \textit{MostPopular} tag recommendation algorithm. Additionally, the participants were divided into two groups at random: (i) \textit{individual}, where the participants only saw their own resources and tags, and (ii) \textit{collaborative}, where the participants also saw the resources and tags of the other users in the group. Thus, in the \textit{collaborative} setting, the tag recommendations were calculated based on the categorized resources and tags of the other users as well. The outcomes of our online study show that, in the \textit{collaborative} setting, \textit{SoMe} provides significantly higher tag recommendation acceptance rates than the \textit{MostPopular} approach. In the \textit{individual} setting, we do not observe a significant difference between the two approaches in terms of recommendation acceptance.  Therefore, we find that a cognitive-inspired tag recommendation algorithm based on a transparent model of human episodic memory supports users in \textit{collaborative} tagging settings. We have validated these findings in a follow-up paper using a similar tag recommendation approach termed \textit{3Layers}, which we have presented at the \textit{International World Wide Web conference 2018 (TheWebConf)}~\cite{kowald2018impact}. 

\begin{figure}[t!]
   \centering
     \captionsetup[subfigure]{justification=centering}
     \subfloat[][Calculation of \textit{BLL}'s $d$ paramenter\\Linear regression: $\alpha$~=~-1.480]{ 
      \includegraphics[width=0.49\textwidth]{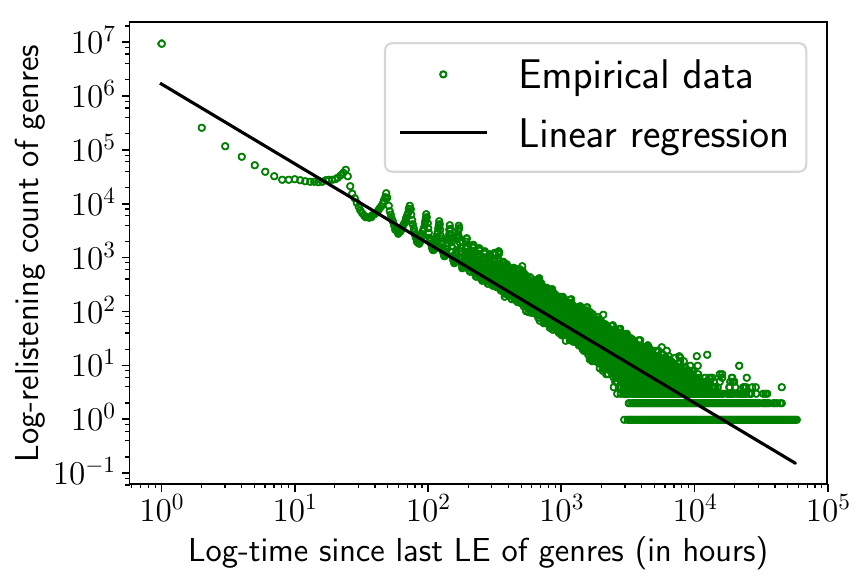} 
   }
     \subfloat[Recall/precision plots for\\$k = 1 \ldots 10$ predicted music genres]{ 
      \includegraphics[width=0.49\textwidth]{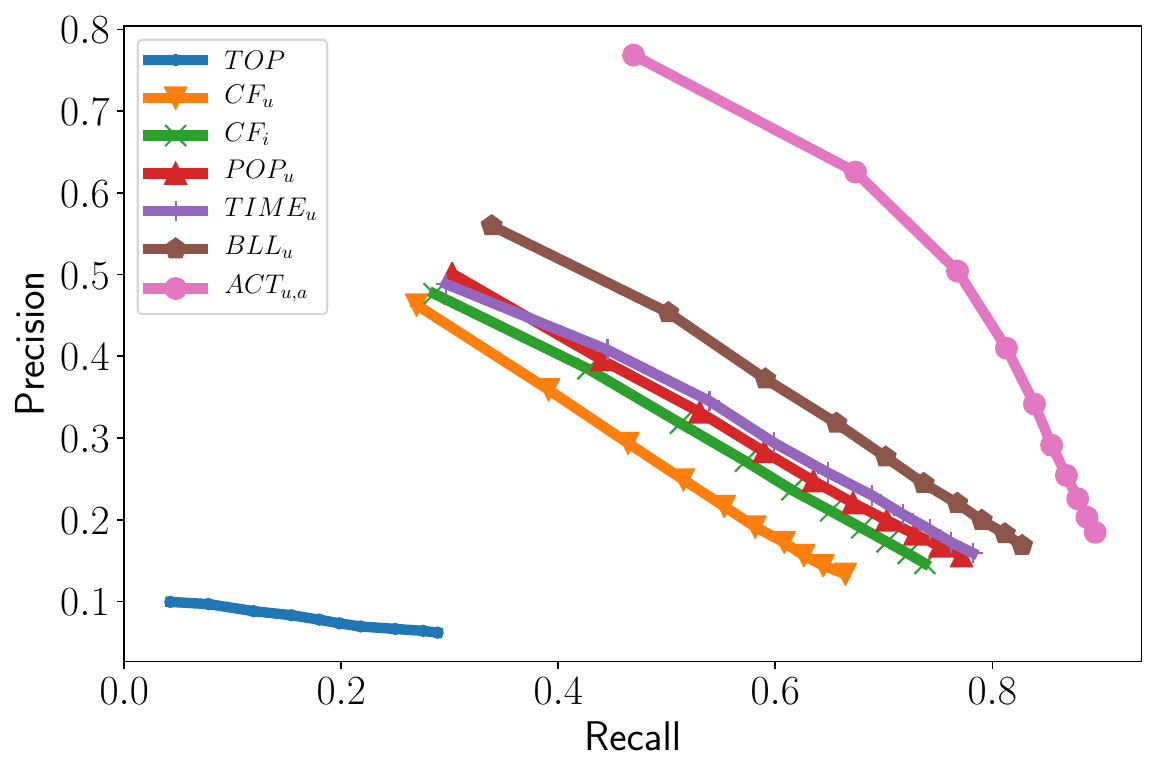}
   }
   \caption{(a) Calculation of the \textit{BLL} equation's $d$ parameter. On a log-log scale, we plot the relistening count of the genres over the time since their last listening event (LE), and set $d$ to the slopes $\alpha$ of the linear regression lines~\cite{lex2020modeling}. (b) Recall/precision plots for $k = 1 \ldots 10$ predicted genres of the baselines, and our $BLL_u$ and $ACT_{u,a}$ approaches. $ACT_{u,a}$ achieves the highest accuracy~\cite{kowald2020utilizing}.}
   \label{fig:actr_results}
\end{figure}

\idTISMIR~and~\idIUI~present the second set of our cognitive-inspired recommendation algorithms based on activation processes in human memory as defined by the cognitive architecture ACT-R~\cite{anderson2004integrated} (see Section \ref{s:rel_cognitive}). We introduce two algorithms for a transparent modeling and prediction approach for music genre preferences of users: (i) $BLL_u$, which implements the \textit{base-level learning} (\textit{BLL}) equation of ACT-R as described in~\idTISMIR, and (ii) $ACT_{u,a}$, which extends $BLL_u$, and implements the full \textit{activation equation} of ACT-R as described in~\idIUI. We evaluate these approaches using dataset samples containing preferences (listening events) of users of the Last.fm music platform, based on the \textit{LFM-1b} dataset~\cite{schedl2016lfm,schedl2017large}. 

Figure~\ref{fig:actr_results} (a) illustrates the impact of time on the re-listening behavior of users in our Last.fm dataset sample. We find that users tend to listen to music genres to which they have listened to very recently, and that this temporal decay follows a power-law distribution as suggested by the \textit{BLL} equation of ACT-R~\cite{anderson1991reflections}. We use the slope $\alpha$ of the linear regression of this data to set \textit{BLL}'s $d$ parameter. Figure~\ref{fig:actr_results} (b) shows the accuracy of our approaches compared to five baseline algorithms: $TOP$ suggests the most popular genres in the system, $CF_u$ and $CF_i$ represent user-based and item-based CF, and $POP_u$ and $TIME_u$ suggest the most popular and most recent genres listened to by $u$, respectively. We find that $BLL_u$ outperforms all baselines, and that $ACT_{u,a}$ outperforms $BLL_u$ by also taking into account the current context of music listening (i.e., the genres of the artist $a$ to which the user $u$ listened to most recently) via the spreading activation component. Our findings show the usefulness of activation processes in human memory for a transparent design process of music recommendation algorithms, which also leads to high recommendation accuracy. We have validated these findings for the task of hashtag recommendations~\cite{kowald2017temporal,info_hashtag_2019}, and for the task of music artist recommendations, which we have presented at the \textit{International Society for Music Information Retrieval (ISMIR)} conference 2019~\cite{ismir_lfm_2019}. 

Finally, \idFNT~surveys and categorizes recommender systems that draw on psychological theories for a transparent design, implementation, or evaluation process of recommendations. We term this type of recommender systems \textit{psychology-informed recommender systems} and categorize them into three groups: (i) cognition (or cognitive)-inspired, (ii) personality-aware, and (iii) affect-aware recommender systems (see Section~\ref{s:rel_psych}). We also discuss open issues in this research field, for example, the need to incorporate psychological considerations into the design process of user-centric recommender system evaluation studies.

\subsubsection{Contribution 2: Illustrating to What Extent Components of the Cognitive Model ACT-R Contribute to Recommendations (2022-2023)}

In~\idRecSysACTR, we discuss transparency aspects of music recommendations generated via ACT-R by illustrating to what extent components of ACT-R have contributed to the generation of recommendation lists. We investigate three ACT-R components described in Section~\ref{s:rel_cognitive}: (i) \textit{the base level learning} (\textit{BLL}) equation, which describes the ``current obsession'' of a user (i.e., frequently and recently listened tracks), (ii) the spreading activation (S) component, which describes ``current vibes'' of a user (i.e., tracks that are similar to the user's most recently listened track), and (iii) the valuation (V) component, which accounts for ``evergreens'' of the user (i.e., the user's most frequently listened tracks, independent of the recency component). Additionally, we analyze a social component (SC) to account for track recommendations ``from similar listeners'' in the form of user-based CF.

\begin{figure}[t!]
    \centering
    \includegraphics[width=0.95\textwidth]{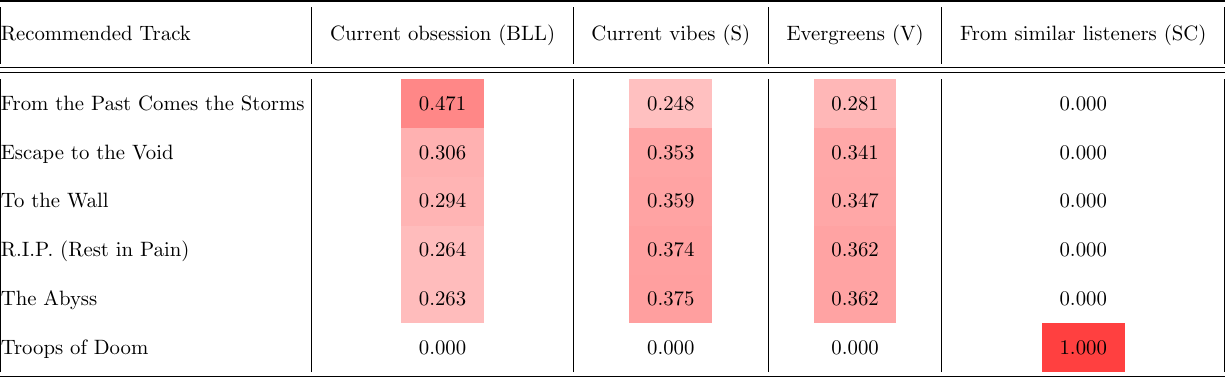}
    \label{fig:explain_recs}
    \caption{Heatmap illustrating the relative contribution of three ACT-R components (BLL, S, and V) and one social component (SC) to the recommendation scores of six recommended tracks for a randomly chosen Last.fm user~\cite{recsys_actr_2023}.}
    \label{fig:explain_table}
\end{figure}

Figure~\ref{fig:explain_table} shows six recommended tracks for a randomly selected user in our newly created Last.fm dataset sample~\cite{actr_dataset_2023_7923900} based on the \textit{LFM-2b} dataset~\cite{melchiorre2021investigating,schedl2022lfm}. The heatmap illustrates how the music track recommendations are calculated by showing the relative contribution of these four components to the recommendation score of a track. We see that the components contribute differently to the recommended tracks. For example, for the first track ``From the Past Comes the Storms'', the current obsession (\textit{BLL}) of the user is most important, while for the last track ``Troops of Doom'' solely the social component (SC) contributes to the recommendation calculation. Based on this, concrete explanations could be derived for all recommendations generated with this model. For example, ``this track was recommended because of your \textit{current obsession}'', or ``this track was recommended because of \textit{similar listeners}''. We discuss transparency aspects of additional components of ACT-R for music recommendations in a chapter for the ``\textit{A Human-centered Perspective of Intelligent Personalized Environments and Systems}'' Springer book, which was recently accepted for publication~\cite{humanizebook}. 

\section{Privacy and Limited Preference Information in Recommender Systems}
\label{sec:cont_privacy}

Limited availability of user preference information (e.g., clicks) could be one consequence of data protection initiatives or of the users' privacy concerns in recommender systems~\cite{knijnenburg2013making,larson2017towards,valdez2019users,biega2020operationalizing,mehdy2021privacy} (e.g., users are not willing to share preferences, or to sign in to the system). Thus, we discuss the findings of two studies that address the limited availability of user preference information in the settings of session-based and cold-start recommendations~\idRecSysTRUST~\idUMUAI~(\textit{Contribution 3}). Additionally, we address varying privacy constraints of users in recommender systems (e.g., hiding preferences)~\idEcirMETA, and the accuracy-privacy trade-off of differentially-private recommender systems~\idTIST. Finally, we survey and categorize the literature on \textit{differential privacy} in collaborative filtering~\idFRONTPRI~(\textit{Contribution 4}).
\vspace{-1mm}
\subsubsection{Contribution 3: Addressing Limited User Preference Information in Cold-Start and Session-based Recommendation Settings (2018-2020)}

\idRecSysTRUST~presents a trust-based CF approach for addressing the user cold-start problem in recommender systems (see Section~\ref{s:rel_sparse_cold}). Specifically, we aim to exploit implicit and explicit connections between users in trust networks~\cite{massa2007trust} to find the $k$ nearest neighbors and to overcome the limited availability of user preference information in this setting. By employing the idea of \textit{regular equivalence} via \textit{Katz} similarity~\cite{helic2014regular}, we do not only find neighbors that share the \textit{same} trust connections, but also neighbors that have \textit{similar} trust connections (i.e., neighbors with similar structural roles in the network). We evaluate our approach using a dataset from the consumer reviewing portal Epinions~\cite{massa2007trust}, which allows users to specify trust connections to other users. We find that our approach outperforms related approaches (e.g., based on Jaccard similarity~\cite{chia2011exploring}) in terms of recommendation accuracy for cold-start users. In our follow-up work~\cite{duricic2020empirical}, we employ graph embedding techniques on the trust network of users by evaluating graph embedding methods such as \textit{graph factorization}~\cite{ahmed2013distributed}, \textit{DeepWalk}~\cite{perozzi2014deepwalk}, or \textit{Node2Vec}~\cite{grover2016node2vec} for the user cold-start problem. We find that \textit{Node2Vec} and \textit{DeepWalk} provide the highest recommendation accuracy and user coverage~\cite{ge2010beyond} across all methods.

\idUMUAI~presents our research on using variational autoencoders for session-based job recommendations. Specifically, to provide personalized job recommendations to users in a setting, in which we do not have full user preference histories available, we employ autoencoders to create latent representations of the limited preference information available in the anonymous user sessions (see Section~\ref{s:rel_sparse_cold}). Our approach recommends jobs within new sessions by employing a $k$-nearest neighbor approach based on the inferred latent session representations generated via standard autoencoders~\cite{bengio2006greedy}, denoising autoencoders~\cite{vincent2008extracting}, and variational autoencoders~\cite{kingma2014auto}. Our evaluation results on session-based job recommendation datasets (e.g., based on XING from the RecSys challenge 2017~\cite{abel2017recsys}) show that our approach based on variational autoencoders provides the most robust results compared to state-of-the-art methods such as \textit{GRU4Rec}~\cite{hidasi2018recurrent}, \textit{session-KNN}~\cite{jannach2017recurrent}, or \textit{sequential session-KNN}~\cite{ludewig2018evaluation}. Here, we do not only evaluate recommendation accuracy, but also novelty metrics~\cite{vargas2011rank} such as system-based novelty (i.e., how unexplored is the recommended job in general~\cite{pu2011user}) and session-based novelty (i.e., how surprising is the recommended job for the current user session~\cite{zhang2012auralist}). To further illustrate the usefulness of variational autoencoders for recommendations, in another paper~\cite{schedl2021listener}, we utilize them to incorporate a user's country information into context-aware music recommendations. Specifically, we incorporate the users' country context into the variational autoencoder architecture via a gating mechanism. Our evaluation results show that our country- and context-aware recommendation approach provides higher recommendation accuracy than related baselines (e.g., variational autoencoders without country information~\cite{liang2018variational}).

\subsubsection{Contribution 4: Addressing Users' Privacy Constraints and the Trade-Off Between Accuracy and Privacy in Recommendations (2021-2023)}

\idEcirMETA~studies the robustness of meta matrix factorization (\textit{MetaMF}) against privacy constraints of users in recommender systems. For this, we conduct a reproducibility study of the original \textit{MetaMF} paper~\cite{lin2020meta}, and investigate the sensitivity of this approach to the limited availability of user preference information, e.g., when users employ privacy constraints by hiding a certain part of their preferences from the system (see Section~\ref{s:rel_privacy}). Therefore, we deactivate the \textit{meta learning}~\cite{vanschoren2019meta} component to evaluate the robustness of \textit{MetaMF} against varying privacy constraints. Additionally, we study how users that differ in their profile size (i.e., number of ratings or implicit item preferences) are affected by varying privacy constraints. On the five datasets \textit{Douban}~\cite{hu2014your}, \textit{Hetrec-MovieLens}~\cite{cantador2011second}, \textit{MovieLens} 1M~\cite{harper2015movielens}, \textit{Ciao}~\cite{guo2014etaf}, and \textit{Jester}~\cite{goldberg2001eigentaste} (we share the dataset samples via \textit{Zenodo}~\cite{mullner_peter_2020_4031011}), we demonstrate that \textit{meta learning} is essential for \textit{MetaMF}'s robustness against users' privacy constraints. We also show that users with small profiles can afford a higher degree of privacy than users with large profiles.

\begin{figure}[t!]
    \centering
    \includegraphics[width=1.0\linewidth]{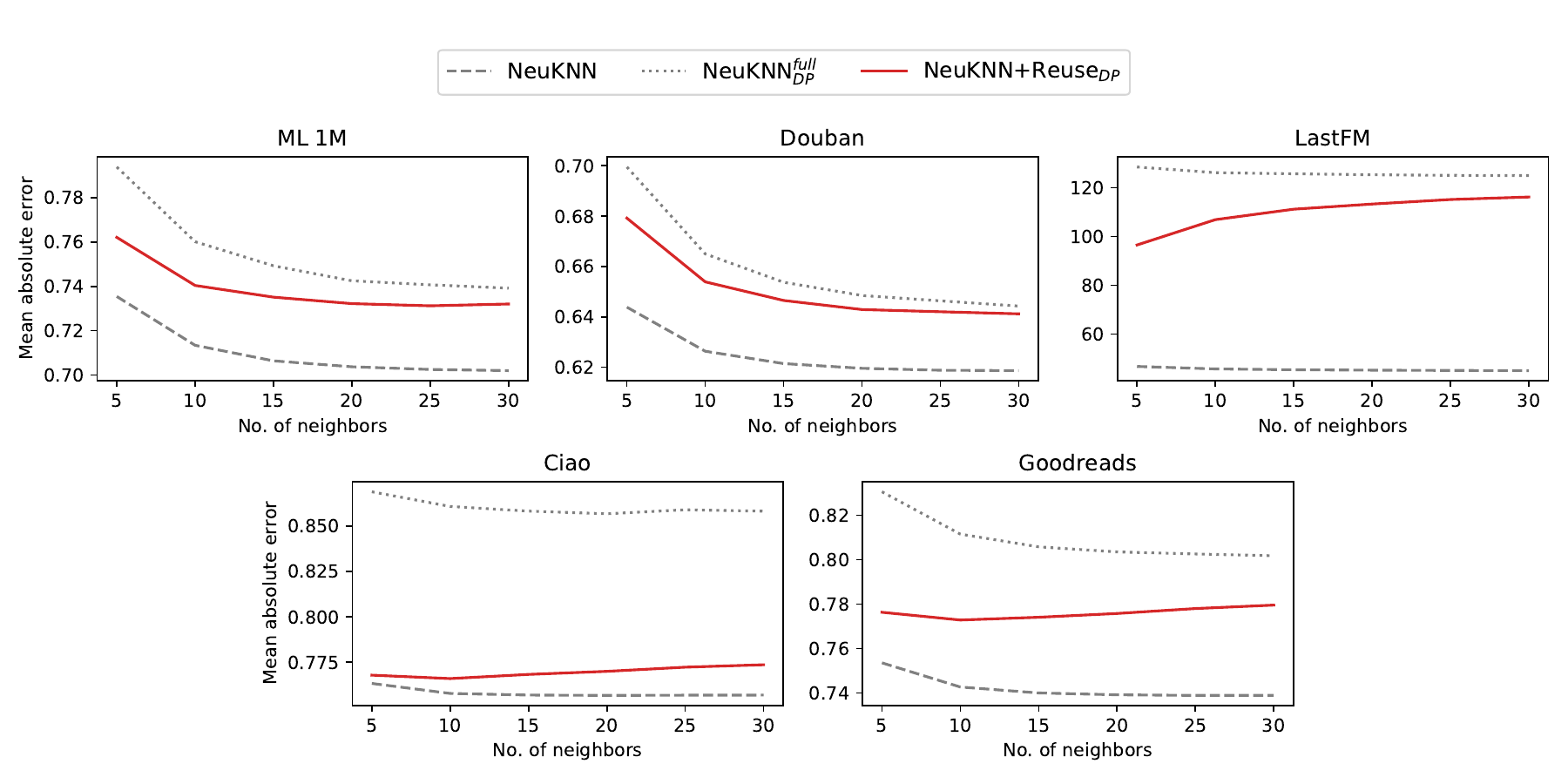}
    \caption{
    Mean absolute error (MAE) of neural-based \emph{KNN} recommender system variants.
    Our results indicate that combining neighborhood reuse with \textit{differential privacy} (\emph{NeuKNN+Reuse}$_{DP}$) yields better accuracy (lower MAE) than neural-based methods that do not apply neighborhood reuse (\emph{NeuKNN}$^{full}_{DP}$)~\cite{tist_dp_2023}. 
    }
    \label{fig:reuseknn_accuracy}
\end{figure} 

\idTIST~addresses the accuracy-privacy trade-off in differentially-private recommender systems. Specifically, we propose our \emph{ReuseKNN} recommendation approach, which aims to reduce the decrease in accuracy due to the application of \textit{differential privacy}~\cite{dwork2008differential,dwork2014algorithmic} on users' preference information~\cite{berkovsky2012impact}. We achieve this by identifying small but highly reusable neighborhoods for $k$-nearest neighbor-based recommendation approaches. Therefore, only this small set of users needs to be protected with \textit{differential privacy}, and the majority of the users do not need to be protected, since they are rarely exploited as neighbors, i.e., they have a small privacy risk~\cite{liu2010framework} as defined in Section~\ref{s:rel_privacy}. We find that with \emph{ReuseKNN}, in the case of a Last.fm dataset sample, only 68.20\% of the users need to be protected with \textit{differential privacy}, while a traditional \emph{UserKNN} approach~\cite{herlocker1999algorithmic} requires the protection of 99.89\% of the users. We validate if this also leads to an improved accuracy-privacy trade-off in various recommendation settings. Figure~\ref{fig:reuseknn_accuracy} shows the recommendation accuracy results of neural-based CF approaches~\cite{he2017neural} when our neighborhood reuse concept is applied (\emph{NeuKNN+Reuse}$_{DP}$, i.e., only vulnerable users are protected), and when it is not applied (\emph{NeuKNN}$^{full}_{DP}$, i.e., all users are protected). Additionally, we include a baseline approach without any application of \textit{differential privacy} (\emph{NeuKNN}, i.e., no users are protected). We see that (i) \emph{NeuKNN} provides the best accuracy results according to the mean absolute error~\cite{willmott2005advantages}, but without any privacy guarantees, (ii) \emph{NeuKNN}$^{full}_{DP}$ provides the worst accuracy results, but with the highest privacy guarantees, and (iii) that our \emph{NeuKNN+Reuse}$_{DP}$ approach provides a better accuracy-privacy trade-off than the other methods. Additionally, in this work, we outline connections between privacy, and item coverage~\cite{herlocker2004evaluating}, popularity bias~\cite{ahanger2022popularity,mullner2024impact}, and fairness~\cite{ekstrand2018privacy}.

Finally, \idFRONTPRI~further discusses the accuracy-privacy trade-off in differentially-private recommendations by surveying the literature in this field. Therefore, we identify 26 papers that apply \textit{differential privacy} either (i) to the user representations (e.g., as we do it in~\idTIST), (ii) directly to the recommendation model updates (e.g., when calculating gradients locally), or (iii) after the recommendation model training process (e.g., applying noise to the trained user and item embeddings). We find that these papers address the accuracy-privacy trade-off in three different ways: (i) using auxiliary data to foster recommendation accuracy (e.g., incorporate preferences of other users), (ii) reducing the noise level that is needed (e.g., requiring the minimal amount of noise to still ensure \textit{differential privacy}), and (iii) limit when to apply \textit{differential privacy} (e.g., as we do it in~\idTIST). 

\section{Fairness and Popularity Bias in Recommender Systems}
\label{sec:cont_fairness}

This section discusses our research on fairness and popularity bias in recommender systems. This contains four publications that study popularity bias for user groups that differ in mainstreaminess (i.e., users' inclination towards mainstream content~\cite{bauer2019global}) and gender~\idEcirPOP~\idEPJ~\idRecSysLBR~\idBiasMEDIA~(\textit{Contribution 5}). This section also describes two papers on understanding popularity bias mitigation and amplification using online and offline evaluation studies~\idEcirPRESSE~\idBiasCALIBRATION~(\textit{Contribution 6}). Another journal article analyzes the long-term dynamics of fairness (e.g., \textit{individual} vs. \textit{group fairness} trade-offs) in algorithmic decision support in a labor market setting using agent-based modeling techniques~\idSCIREP~(\textit{Contribution 7}).

\subsubsection{Contribution 5: Measuring Popularity Bias for User Groups Differing in Mainstreaminess and Gender (2020-2022)}

\begin{figure}[t!]
\centering
   \subfloat[MostPopular]{
      \includegraphics[width=.33\textwidth]{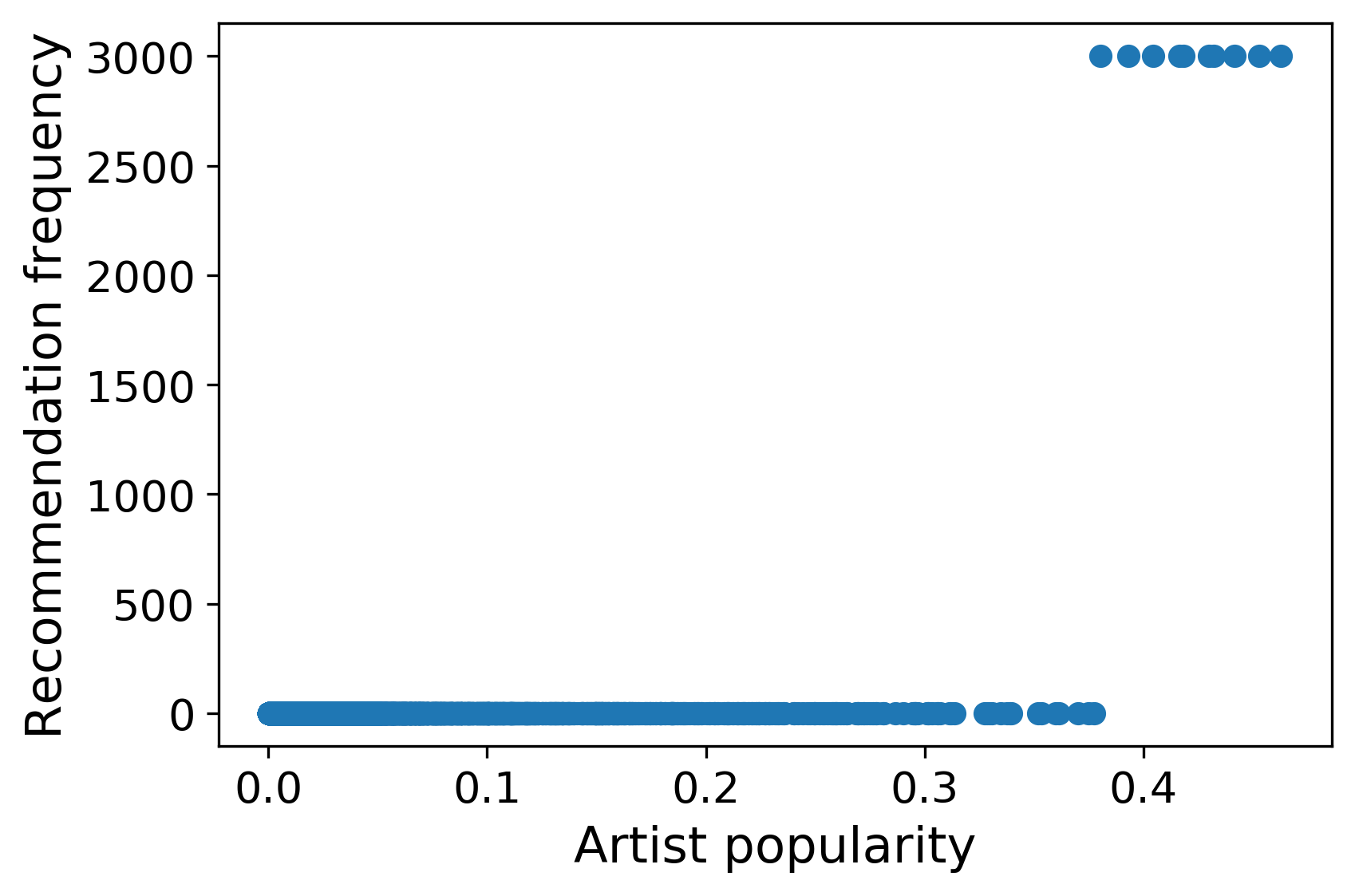}}
   \subfloat[UserKNN]{
      \includegraphics[width=.33\textwidth]{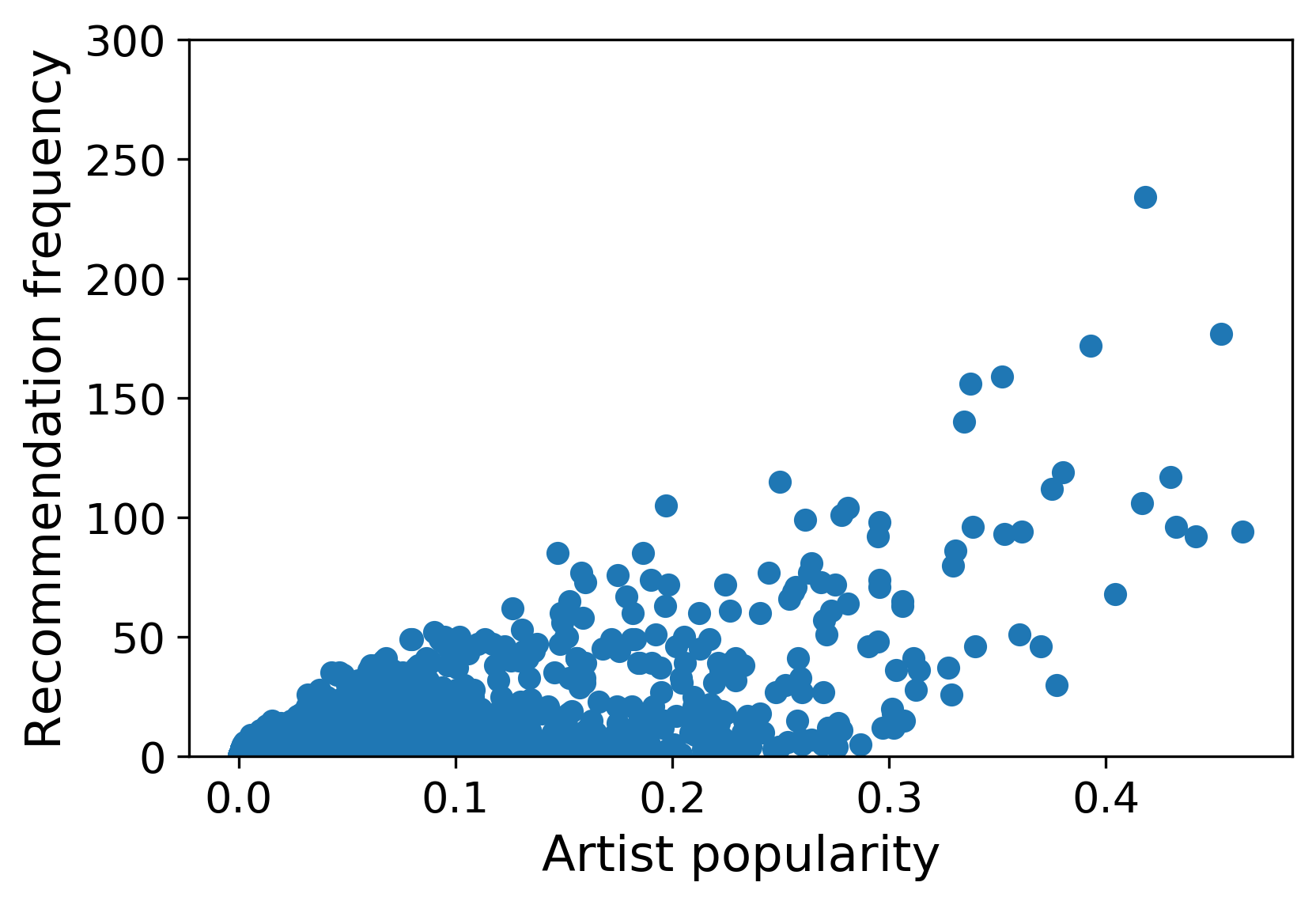}}
   \subfloat[NMF]{
      \includegraphics[width=.33\textwidth]{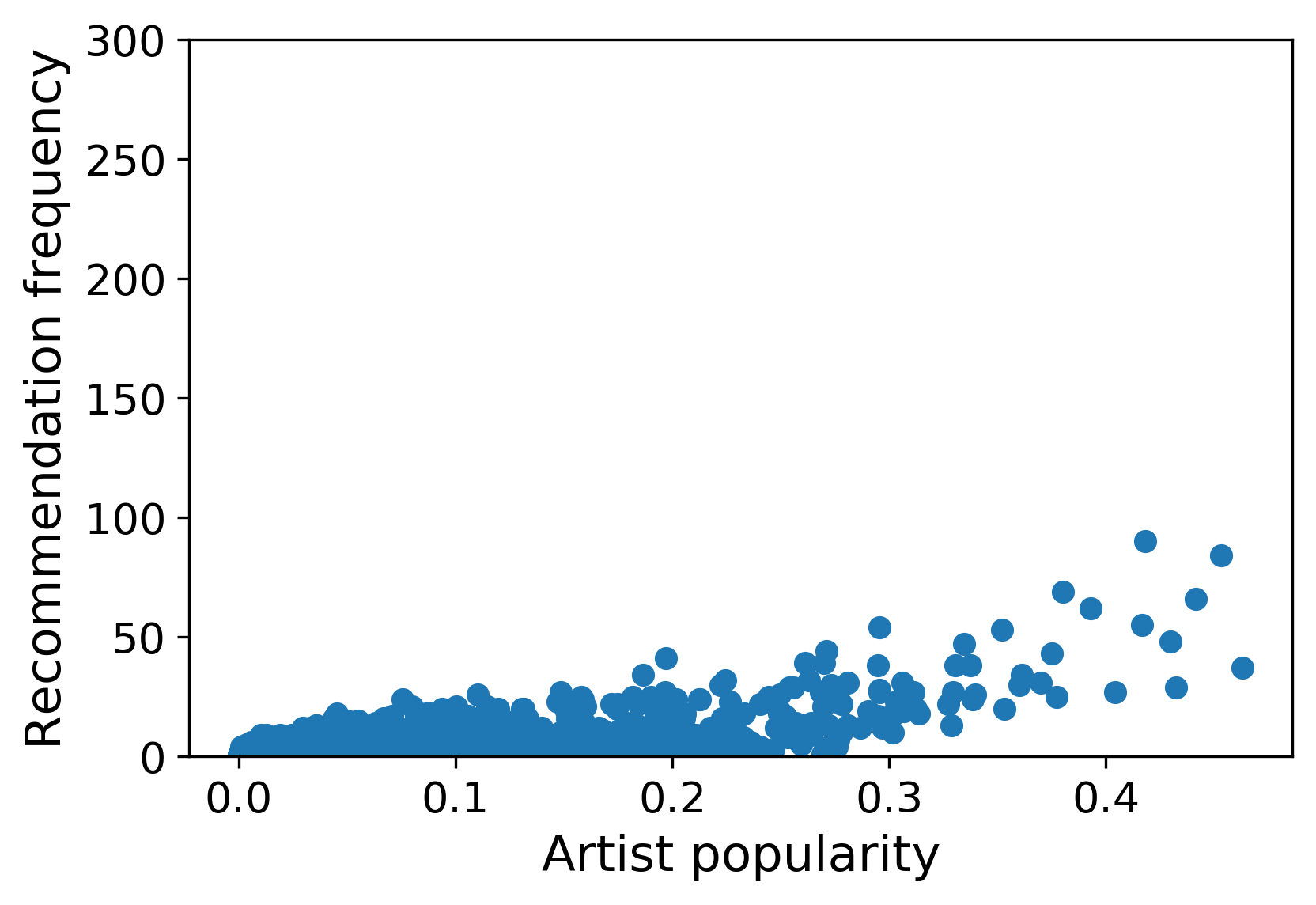}}
   \caption{Correlation of music artist popularity and recommendation frequency. All three algorithms investigated tend to favor popular music artists~\cite{kowald2020unfairness}.}
   \label{fig:rec_popularity}
\end{figure}

\begin{figure}[t!]
    \centering
   \subfloat[MAE of \textit{MS} and \textit{BeyMS} users.]{
      \includegraphics[height=3.3cm]{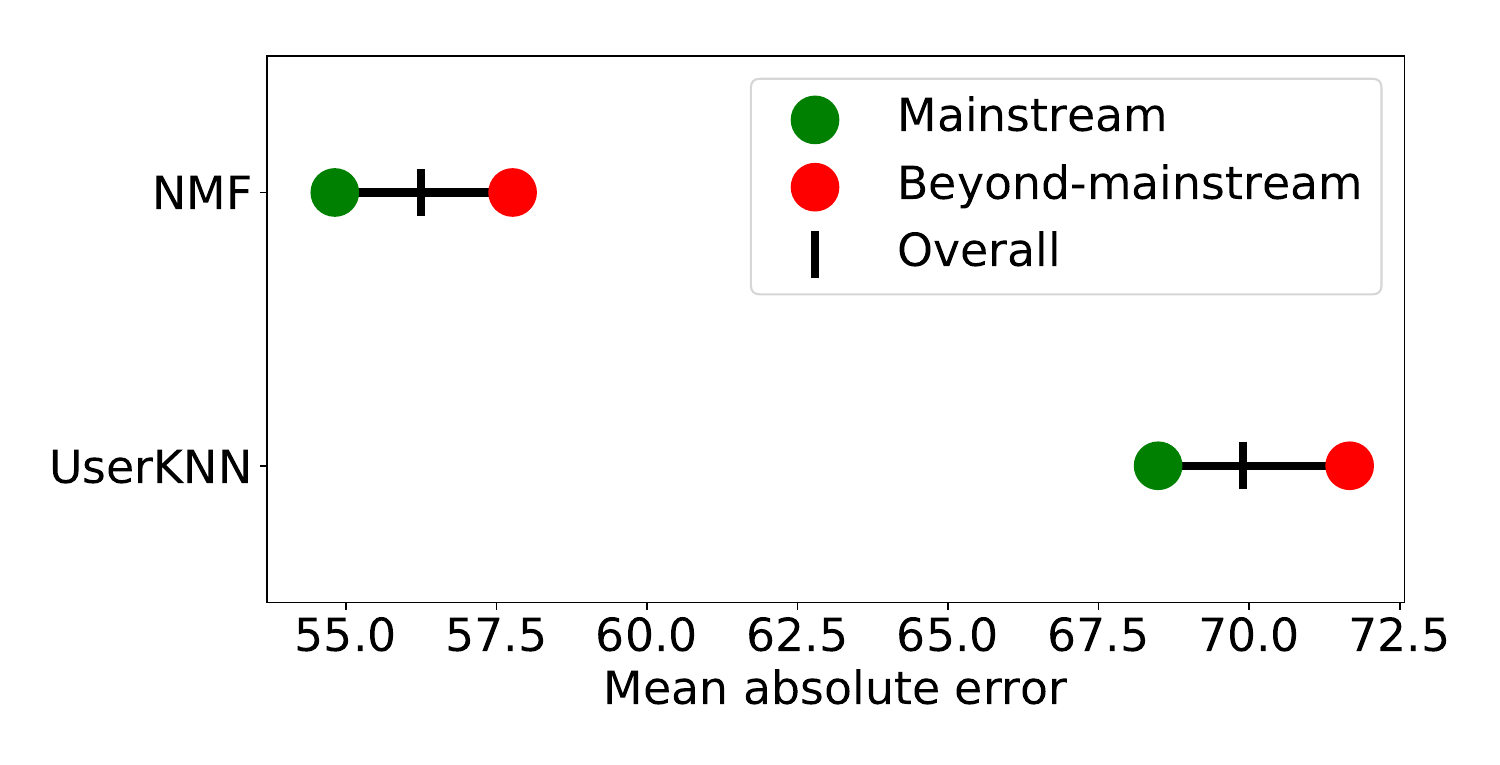}
      \label{fig:mae_beyms_ms}}    
   \subfloat[MAE of \textit{BeyMS} subgroups by \textit{NMF}.]{
      \includegraphics[height=3.3cm]{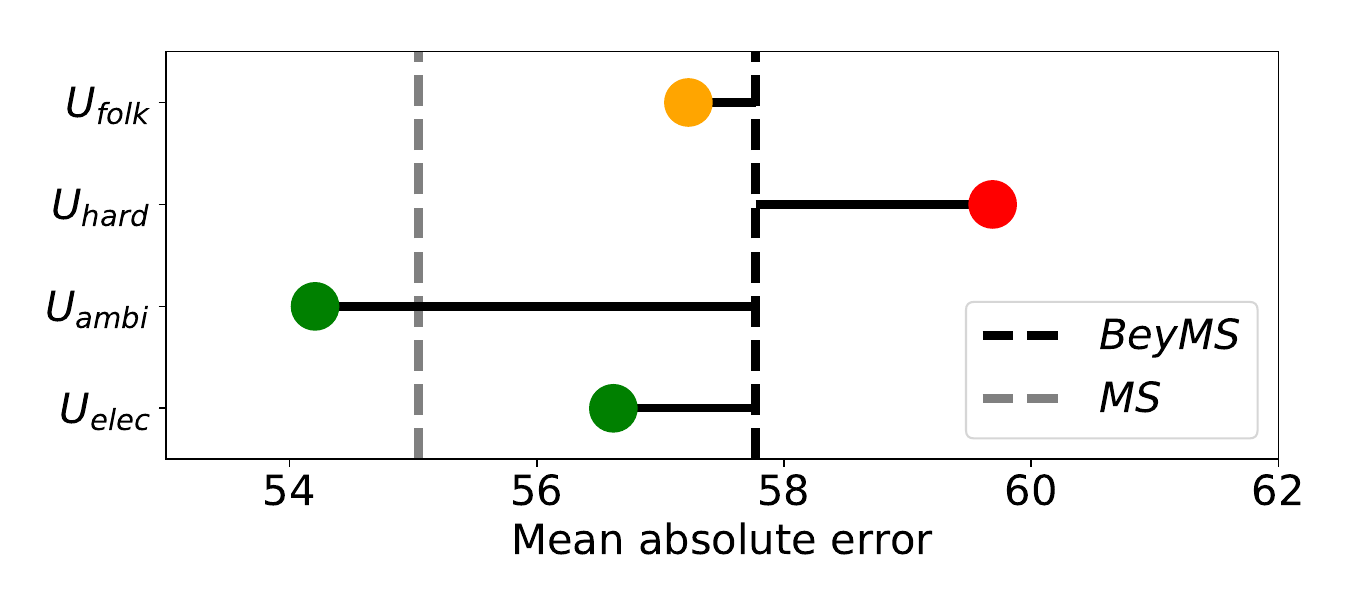}
      \label{fig:mae_subgroups}}  
    \caption{(a) Recommendation accuracy measured by the mean absolute error (MAE) of \textit{NMF} and \textit{UserKNN} for mainstream (\textit{MS}) and beyond-mainstream (\textit{BeyMS}) user groups in Last.fm: \textit{BeyMS} users receive a substantially lower recommendation quality (i.e., higher MAE) compared to \textit{MS} users. 
    (b) Comparison of the MAE scores reached by \textit{NMF} for the four \textit{BeyMS} subgroups with the ones reached by \textit{NMF} for \emph{BeyMS} (black dashed line) and \emph{MS} (grey dashed line). There are substantial differences between the subgroups in terms of MAE, especially when comparing $U_{hard}$ with $U_{ambi}$, i.e., two subgroups differing in their openness to music listened to by users of other subgroups~\cite{kowald2021support}.}
     \label{fig:motivation_example}
\end{figure}

\idEcirPOP~analyzes the unfairness of popularity bias in music recommendations. Specifically, we reproduce a study by Abdollahpouri et al.~\cite{abdollahpouri2019unfairness}, in which the authors find that personalized recommendation algorithms in the movie domain are biased towards popular items, and that this popularity bias also leads to the unfair treatment of users with little interest into popular content (see Section~\ref{s:rel_popbias}). We conduct this reproducibility study in the music domain using a newly created dataset sample~\cite{kowald_dominik_2019_3475975} gathered from Last.fm. Figure~\ref{fig:rec_popularity} shows that our results are in line with the ones of~\cite{abdollahpouri2019unfairness} since all evaluated recommendation algorithms tend to favor popular items also in the music domain. In the case of the \textit{MostPopular} algorithm, as expected, the strongest evidence for popularity bias can be found. In the case of traditional \textit{UserKNN}~\cite{herlocker1999algorithmic} and \textit{Non-negative Matrix Factorization} (\textit{NMF})~\cite{luo2014efficient}, we also see a positive relationship between item (i.e., music artist) popularity and recommendation frequency. Finally, for \textit{UserKNN} and \textit{NMF}, we find that beyond-mainstream (\textit{BeyMS}) users receive less accurate recommendations than mainstream (\textit{MS}) users (see Figure~\ref{fig:mae_beyms_ms}).

In~\idEPJ, we analyze the unfairly treated \textit{BeyMS} user group in more detail by identifying subgroups of beyond-mainstream music listeners. For this, we create a new dataset termed \textit{LFM-BeyMS}, which contains (among others) audio features of the music tracks listened to by more than 2,000 \textit{BeyMS} users. Using these audio features and unsupervised clustering techniques, we identify four clusters of beyond-mainstream music and music listeners: (i) $U_{folk}$, listeners of music with high acousticness such as ``folk'', (ii) $U_{hard}$, listeners of high energy music such as ``hardrock'', (iii) $U_{ambi}$, listeners of music with high acousticness and instrumentalness such as ``ambient'', and (iv) $U_{elec}$, listeners of high energy music with high instrumentalness such as ``electronica''. Figure~\ref{fig:mae_subgroups} shows that there is a substantial difference in recommendation accuracy between these subgroups of \textit{BeyMS} users. While $U_{ambi}$ users, on average, even receive better recommendation accuracy results than \textit{MS} users, $U_{hard}$ users receive the worst recommendation accuracy results. 
When relating our results to the openness of the subgroups' users towards music listened to by the other subgroups, we find that $U_{ambi}$ is the most open group, while $U_{hard}$ is the least open group. This is in line with related research~\cite{tintarev2013adapting}, which has shown that a user's openness towards content consumed by other users is positively correlated with recommendation accuracy.

\idRecSysLBR~studies if popularity bias in music recommender systems affect users of different genders in the same way. To answer this question, we analyze seven recommendation algorithms, \textit{Random}, \textit{MostPopular}, \textit{ItemKNN}~\cite{sarwar2001item}, \textit{Sparse Linear Method} (\textit{SLIM})~\cite{ning2011slim}, \textit{Alternating Least Squares Matrix Factorization} (\textit{ALS})~\cite{hu2008collaborative}, \textit{Matrix Factorization with Bayesian Personalized Ranking} (\textit{BPR})~\cite{rendle2009bpr}, and \textit{Variational Autoencoder for CF} (\textit{VAE})~\cite{liang2018variational}, on a Last.fm dataset sample based on the \textit{LFM-2b} dataset~\cite{melchiorre2021investigating,schedl2022lfm}. 
We find that all personalized recommendation algorithms investigated in this study, except for \textit{SLIM}, intensify the popularity bias for female users. Thus, not only user groups differing in mainstreaminess, but also user groups differing in gender are affected differently by popularity bias. 

Finally,~\idBiasMEDIA~validates the findings of~\idEcirPOP~and~\idEPJ~in three additional multimedia domains, namely (i) movies (\textit{MovieLens-1M}~\cite{harper2015movielens}), (ii) books (\textit{BookCrossing}~\cite{ziegler2005improving}), and (iii) animes (\textit{MyAnimeList}~\cite{myanimelist_dataset2018}). For these datasets, we create dataset samples~\cite{kowald_dominik_2022_6123879} with user groups that differ in their inclination to popular and mainstream content, and analyze popularity bias of various CF-based recommendation algorithms on the levels of items and users. On the item level, we find that the probability of an item to be recommended strongly correlates with the popularity of the item. On the user level, we find that users with the least inclination to popular content also receive the worst recommendation quality.  

\subsubsection{Contribution 6: Understanding Popularity Bias Mitigation and Amplification in Recommendations (2022-2023)}

\idEcirPRESSE~presents an online study on popularity bias mitigation (see Section~\ref{s:rel_popbias}) in a news article recommendation setting. To conduct our online study, we collaborate with \textit{DiePresse}, a popular Austrian online news platform, and discuss the introduction of personalized, content-based news article recommendations into the platform as a replacement for unpersonalized \textit{MostPopular} recommendations. Our content-based recommendation algorithm~\cite{lops2010content,de2015semantics} is based on latent representations of news articles using \textit{Latent Dirichlet Allocation (LDA)}~\cite{blei2003latent}. We conducted our online study in a two-week time window (27th of October 2020 to 9th of November 2020), in which we tracked user preferences (i.e., clicks on news articles) of more than one Million anonymous user sessions, and more than 15,000 signed in (subscribed) users of \textit{DiePresse}. Within our two-week online study, also two significant events happened that could influence the reading behavior of users: (i) the COVID-19 lockdown announcements in Austria on the 31st of October 2020, and (ii) the Vienna terror attack on the 2nd of November 2020.

\begin{figure}[t!]
\centering
\subfloat[\textit{Skewness}]{
\includegraphics[width=.49\textwidth]{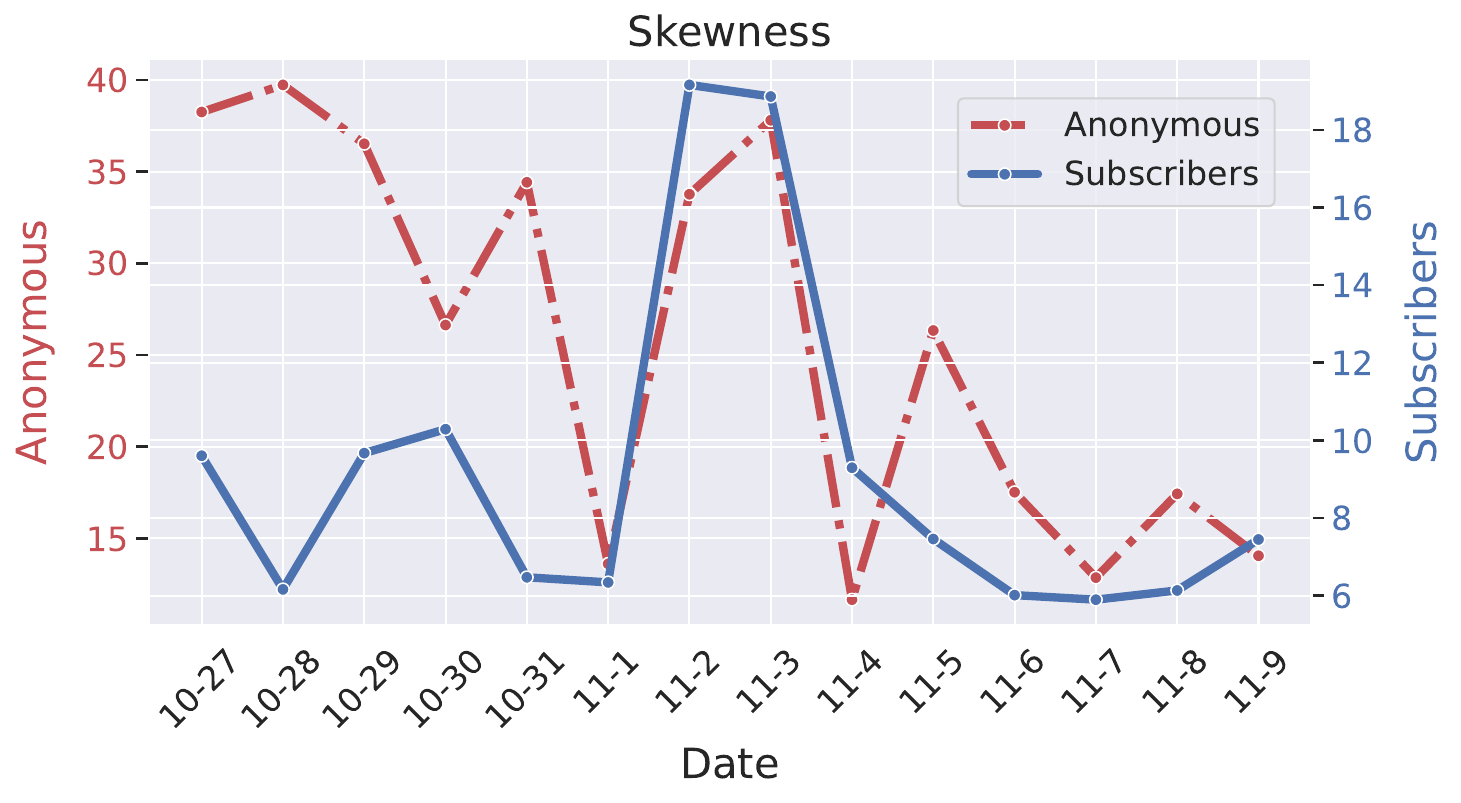}}
~
\subfloat[\textit{Kurtosis}]{
\includegraphics[width=.49\textwidth]{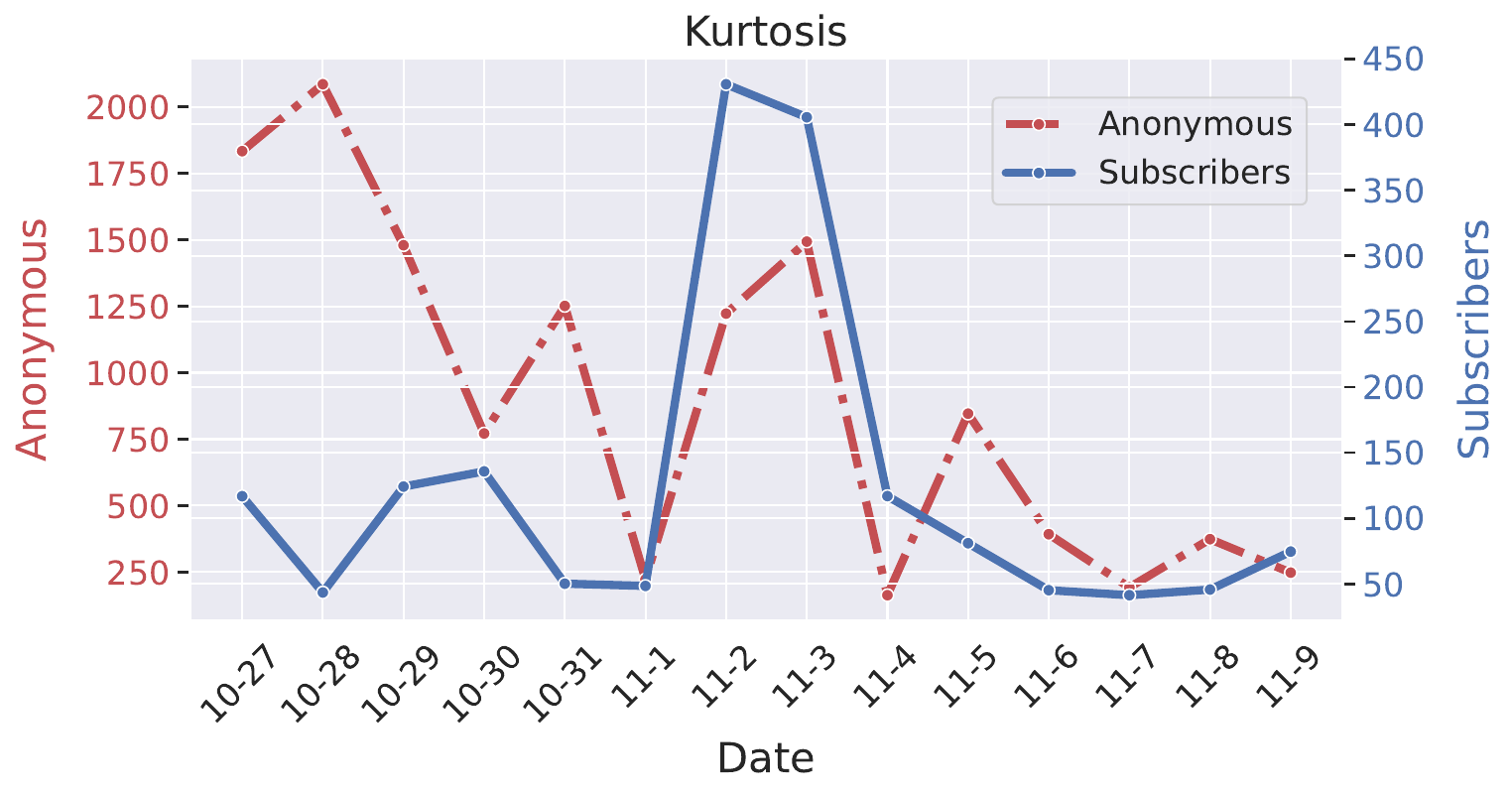}}
\caption{Mitigation of popularity bias in news article consumption, measured by (a) \textit{skewness} and (b) \textit{kurtosis} based on the number of article reads for each day of our two-week online study. At the beginning of the study, the \textit{MostPopular} news article recommendations were replaced by personalized, content-based recommendations. We find that popularity bias can be mitigated by introducing personalized news article recommendations in the case of anonymous users~\cite{lacic2022drives}.}
\label{fig:presse_pop_bias}
\end{figure}

Figure~\ref{fig:presse_pop_bias} shows the results of our online study in terms of \textit{skewness} and \textit{kurtosis} of the news article popularity distribution (i.e., number of article reads) across the two weeks, and for both user groups (i.e., anonymous and subscribed users). Here, \textit{skewness} measures the asymmetry, and \textit{kurtosis} measures the ``tailedness'' of the popularity distribution~\cite{bellogin2017statistical}. For both metrics, high values indicate a popularity biased news consumption, which could lead to filter bubble and echo chamber effects~\cite{flaxman2016filter}. At the beginning of the online study, where \textit{MostPopular} recommendations were shown, we see a large gap between the two user groups: while anonymous users mainly read popular news articles, and thus, are prone to popularity bias, subscribed users show a much more balanced reading behavior. At the end of the study, i.e., after two weeks of personalized recommendations, we see a considerably smaller difference between the two user groups, which means that the introduction of personalized, content-based news article recommendations helped to mitigate popularity bias in the case of anonymous users already after two weeks. However, in the case of significant events, e.g., the Vienna terror attack on the 2nd of November 2020, both user groups are mostly interested into popular articles reporting on the particular event. In another work~\cite{conext_dataeco_2022}, we also find that content-based recommendations can help to mitigate popularity bias in the case of recommendations provided in a data and algorithm sharing platform.

In~\idBiasCALIBRATION, we analyze miscalibration~\cite{steck2018,lin2020} and popularity bias amplification (in terms of the popularity lift metric~\cite{abdollahpouri2019unfairness,abdollahpouri2020connection}) in music, movie, and anime recommender systems. For this, we extend the \textit{MovieLens 1M}~\cite{harper2015movielens}, \textit{LFM-1b}~\cite{schedl2016lfm,schedl2017large}, and \textit{MyAnimeList}~\cite{myanimelist_dataset2018} datasets with genre information of the items, and publish these new dataset samples via \textit{Zenodo}~\cite{dominik_kowald_2022_7428435}. Then we measure accuracy, miscalibration, and popularity bias amplification (i.e., popularity lift) for various recommendation algorithms (e.g., \textit{NMF}~\cite{luo2014efficient} and \textit{co-clustering}-based CF~\cite{george2005scalable}), and for user groups differing in their inclination to popular and mainstream content, i.e., (i) \textit{LowPop} (low interest in popular content), (ii) \textit{MedPop} (medium interest in popular content), and (iii) \textit{HighPop} (high interest in popular content). We find that there is a connection between these three metrics, since the \textit{LowPop} user group, which receives the worst recommendation accuracy results, is also the user group, which receives the most miscalibrated and popularity biased recommendations. 
Finally, we investigate to what extent particular genres contribute to the inconsistency of recommendation performance in terms of miscalibration and popularity bias amplification. We find that there are indeed genres that highly contribute to inconsistent and popularity biased recommendation results. One example is the ``Hentai'' genre in the case of our \textit{MyAnimeList} dataset sample: this is a genre, which is highly popular for a specific user group (i.e., \textit{LowPop}), and unpopular for the other user groups (i.e., \textit{MedPop} and \textit{HighPop}).

\subsubsection{Contribution 7: Studying Long-Term Dynamics of Fairness in Algorithmic Decision Support (2022-2023)}

\idSCIREP~studies the long-term dynamics of fairness in algorithmic decision support (see Section~\ref{s:rel_fairness}) in a labor market setting~\cite{gachter2002fairness}. Specifically, we develop and evaluate an agent-based simulation model to investigate the impact of decisions caused by a public employment service that decides which jobseekers receive targeted help using a decision support tool. This tool uses a logistic regression model~\cite{wright1995logistic} to classify jobseekers into low- and high-prospects. We use synthetic data that describes a pool of jobseekers with unevenly distributed skills between two groups that differ with respect to a protected attribute. We test two variants of our prediction model: (i) a biased version that augments knowledge about the actual skills of a jobseeker with knowledge about the protected attribute, and (ii) an unbiased version that solely relies on the skills of a jobseeker. Based on the classification into low-prospects and high-prospects, our agent-based simulation model updates the skills of the jobseekers after each iteration accordingly (e.g., a high-prospect receives help, and thus also the skills of this jobseeker increase).

Our results show that there is a trade-off between different long-term fairness goals. On the one hand, when using the biased prediction model, the inequality between the two protected groups is reduced at the end of the simulation. This means, that \textit{statistical parity} in the dataset~\cite{barocas2016big} increases, and that the system is fair from a \textit{group fairness} perspective. However, on the other hand, the number of misclassifications of jobseekers in the unprivileged group increases: some jobseekers are classified as low-prospect mainly because of their sensitive attribute, although they should belong to the high-prospect group. This means that the system is unfair from an \textit{individual fairness} perspective. Although this study was not conducted in the field of recommender systems, we believe that the applied method (i.e., agent-based modeling) could also be of use when studying long-term fairness dynamics of recommender systems. Additionally, our findings with respect to the trade-off between \textit{individual} and \textit{group fairness} are also highly relevant for the research area of fair recommender systems. 

\section{Summary of Contributions and Reproducibility of Research Results}
\label{sec:cont_summary}

This section summarizes the 7 scientific contributions described in the previous sections. Additionally, the reproducibility of the findings are discussed.
\vspace{-1mm}
\subsubsection{List of Contributions}
\begin{enumerate}
    \item \textbf{Using cognitive models for a transparent design and implementation process of recommender systems (2018-2021)}: we propose a tag recommendation approach based on a model of human episodic memory~\idHCI, and two music recommendation approaches based on activation process in human memory~\idTISMIR~\idIUI. Additionally, we identify three types of psychology-informed recommender systems: (i) cognition-inspired, (ii) personality-aware, and (iii) affect-aware recommender systems~\idFNT.
    \item \textbf{Illustrating to what extent components of the cognitive model ACT-R contribute to recommendations (2022-2023)}: we illustrate to what extent components of ACT-R (e.g., \textit{BLL} or \textit{valuation}) have contributed to the generation of music recommendation lists. Based on this, explanations for the music recommendation could be derived~\idRecSysACTR.
    \item \textbf{Addressing limited user preference information in cold-start and session-based recommendation settings (2018-2020)}: we model a user’s trust network using regular equivalence to address the user cold-start problem~\idRecSysTRUST. Additionally, we demonstrate the usefulness of variational autoencoders for session-based job recommendations~\idUMUAI. 
    \item \textbf{Addressing users' privacy constraints and the accuracy privacy trade-off in recommendations (2021-2023)}: we study privacy constraints of users (e.g., hiding preferences) in meta matrix factorization~\idEcirMETA, design a neighborhood reuse approach~\idTIST, and survey the literature for differentially-private collaborative filtering recommender systems~\idFRONTPRI. 
    \item \textbf{Measuring popularity bias for user groups differing in mainstreaminess and gender (2020-2022)}: we study popularity bias~\idEcirPOP, characteristics of beyond-mainstream users~\idEPJ, and differences with respect to users' gender in music recommendations~\idRecSysLBR. We also show the presence of popularity bias in movie, book, and anime recommendations~\idBiasMEDIA.
    \item \textbf{Understanding popularity bias mitigation and amplification in recommendations (2022-2023)}: we analyze and mitigate popularity in news article recommender systems~\idEcirPRESSE, and study to what extent recommendations amplify popularity bias in the music, movie, and anime domains~\idBiasCALIBRATION.
    \item \textbf{Studying long-term dynamics of fairness in algorithmic decision support (2022-2023)}: we show the usefulness of agent-based modeling techniques for studying long-term dynamics of algorithmic fairness in a labor market setting. Additionally, we find evidence for the presence of the trade-off between \textit{individual} and \textit{group fairness} in this setting~\idSCIREP.
\end{enumerate}

\subsubsection{Reproducibility of Research Results}
To foster the reproducibility of these research results and findings, we provide information on the used source-code and dataset samples in all publications. In cases, in which we create new dataset samples or implement novel recommendation pipelines, we make them freely available via \textit{Zenodo} or \textit{GitHub}. For example, to implement and evaluate our cognitive-inspired recommendation approaches, we build upon our \textit{TagRec} framework~\cite{kowald2014tagrec,trattner2015tagrec,kowald2017tagrec}, and extend it with music recommendation approaches. Another example is our \textit{Last.fm user group} dataset sample~\cite{kowald_dominik_2019_3475975} that can be used to study fairness and popularity bias in recommender systems. Additionally, we contribute to reproducibility studies by presenting two papers enlisted in this habilitation in the reproducibility track of the \textit{European Conference on Information Retrieval (ECIR'2020 and ECIR'2021)}~\idEcirMETA~\idEcirPOP.

A list of the new dataset samples and recommendation pipelines created in the publications that are part of this habilitation is given in the following:

\begin{enumerate}
    \item The \textit{TagRec} framework~\cite{kowald2014tagrec,trattner2015tagrec,kowald2017tagrec} used to design, develop, and evaluate cognitive-inspired algorithms for tag and music recommendations: \url{https://github.com/learning-layers/TagRec}.
    \item A \textit{GitHub} repository with the material to generate sequential music recommendations and to illustrate to what extent the components of the cognitive model ACT-R contribute to the generation of music recommendation lists~\cite{recsys_actr_2023}: \url{https://github.com/hcai-mms/actr}.
    \item A dataset sample based on the \textit{LFM-2b} dataset~\cite{melchiorre2021investigating,schedl2022lfm} used to generate and evaluate sequential music recommendations~\cite{actr_dataset_2023_7923900}. This \textit{Zenodo} repository also contains the pre-calculated embeddings for the \textit{BPR} approach.
    \item Source-code and dataset references for using variational autoencoders in the setting of session-based job recommendations~\cite{lacic2020using}: \url{https://github.com/lacic/session-knn-ae}. This \textit{GitHub} repository also contains implementations of beyond-accuracy evaluation metrics (e.g., diversity and novelty) for session-based recommender systems.
    \item A dataset for studying privacy constraints of different users groups using meta matrix factorization~\cite{mullner_peter_2020_4031011} accompanied by a \textit{GitHub} repository: \url{https://github.com/pmuellner/RobustnessOfMetaMF}.
    \item The material for the differentially-private \textit{ReuseKNN}~\cite{tist_dp_2023} recommender system: \url{https://github.com/pmuellner/ReuseKNN}. This \textit{GitHub} repository also contains the implementation of \textit{Neural CF}, as well as source-code for sampling user preference histories in the datasets.
    \item A dataset for studying beyond-mainstream users in music recommender systems~\cite{peter_mullner_2020_3784765} accompanied by a \textit{GitHub} repository: \url{https://github.com/pmuellner/supporttheunderground}. Apart from popularity bias evaluation metrics, this \textit{GitHub} repository contains implementations of unsupervised clustering techniques to analyze audio features of music tracks. 
    \item Datasets containing different user groups to study fairness and popularity bias in music, movie, book, and anime recommender systems~\cite{kowald_dominik_2019_3475975,kowald_dominik_2022_6123879}. For calculating calibration-based metrics in these settings, an extended version of these datasets also contains genre information for the items~\cite{dominik_kowald_2022_7428435}.
    \item A Python-based pipeline to process the datasets used in~\cite{kowald2020unfairness,kowald2022popularity,ecir_bias_2023} for studying fairness and popularity bias in recommender systems: \url{https://github.com/domkowald/FairRecSys}. This \textit{GitHub} repository can also be used as a basis to develop popularity bias mitigation methods.
\end{enumerate}

By publishing these resources, the author of this habilitation hopes to contribute to reproducible research practices in the field of recommender systems. As mentioned already in Section~\ref{s:rel_main}, the reproducibility of research results is highly important for being able to track progress in recommender systems research.

\newpage


\chapter{Outlook and Future Research}
\label{c:outlook}

This chapter gives an outlook into future research directions of this habilitation.

\subsubsection{Transparency and Cognitive Models in Recommender Systems}

The underlying algorithms of modern recommender systems are often based on purely data-driven machine learning models. Although these approaches provide high accuracy, they are based on principles of artificial intelligence rather than human intelligence. One consequence could be that the logic of these models is not directly understandable by humans, which could lead to non-transparent algorithmic decisions~\cite{sinha2002role}. 
This habilitation has shown that using psychological theories, and modeling the underlying cognitive processes that describe how humans access information in their memory, is one way to overcome this issue, and at the same time, to generate accurate recommendations (see Section~\ref{sec:cont_transparency}).

Besides MINERVA2~\cite{hintzman1984minerva}, the cognitive architecture ACT-R~\cite{anderson2004integrated} provides an excellent basis by formalizing two kinds of human memory: (i) declarative memory, and (ii) procedural memory. The declarative memory corresponds to things that humans know by determining the importance of information chunks, while the procedural memory corresponds to knowledge of how humans do things by defining production rules for making decisions. This habilitation has focused on modeling declarative memory processes for a transparent design process of cognitive-inspired recommender systems. Thus, in future research, I aim to investigate to what extent also the procedural memory module of ACT-R can be used to design recommendation models (e.g., by adapting the \textit{SNIF-ACT}~\cite{fu2007snif} user navigation model). Here, one interesting research question would be if the defined production rules could further contribute to transparency aspects of cognitive-inspired recommender systems. This question could be answered by conducting user studies following well-established procedures in the field (e.g.,~\cite{paramythis2010layered,knijnenburg2012explaining,tintarev2015explaining}).

\subsubsection{Privacy and Limited Preference Information in Recommender Systems}

Privacy is a key requirement for recommender systems, since there are multiple privacy threats to users in these systems. For example, disclosing users' preference information to untrusted third parties~\cite{calandrino2011you}, or inferring users' sensitive attributes such as gender~\cite{zhang2023comprehensive}. Privacy is also related to the issue of limited availability of user preference information, since users increasingly care about their privacy and may not want to share their preferences with the system~\cite{knijnenburg2013making,valdez2019users,mehdy2021privacy}. Additionally, initiatives such as the \textit{European General Data Protection Regulation (GDPR)} restrict the use of user preference information to generate recommendations~\cite{cummings2018role,biega2020operationalizing}. 
This habilitation has addressed session-based and cold-start recommendation settings, and the accuracy-privacy trade-off when applying \textit{differential privacy} to the users' preference information (see Section~\ref{sec:cont_privacy}). 

In the future, I plan to not only study the trade-off between accuracy and privacy, but also to investigate other relevant trade-offs between recommendation objectives. This includes the trade-off between privacy and fairness~\cite{ekstrand2018privacy}. Here, an interesting research question would be if different user groups are treated differently by the accuracy drops due to privacy-preserving technologies, such as \textit{differential privacy}. For this, related studies from the field of private and fair machine learning (e.g.,~\cite{bagdasaryan2019differential}) could be adapted for recommender systems. Additionally, studying privacy dynamics in recommendations using agent-based simulations would be a promising research direction, as described in our position paper~\cite{mullner2021position} presented in the \textit{SimuRec} workshop of \textit{ACM RecSys} 2021.

\subsubsection{Fairness and Popularity Bias in Recommender Systems}

Biases in the perception and behavior of humans are captured, reflected, and potentially amplified in recommender systems~\cite{friedman1996bias,lambrecht2019algorithmic,chen2023bias}. The replication of popularity bias is a common issue in collaborative filtering-based recommender systems, which leads to the overrepresentation of popular items in the recommendation lists. The research presented in this habilitation has shown that users with little interest in popular content receive worse recommendation accuracy than users that like to consume popular content. Based on this, these users are treated in an unfair way by the recommender system (see Section~\ref{sec:cont_fairness}). 

In my future research, I plan to work on popularity bias mitigation methods to reduce the accuracy differences between the user groups, and with this, increase the fairness in the system. For this, not only technical debiasing methods (e.g., in- or post-processing~\cite{abdollahpouri2021user}), but also novel multidisciplinary approaches using models from psychology and physics should be developed. For the former, ACT-R~\cite{anderson2004integrated} could be a promising basis to build strongly personalized user models, and for the latter, techniques from physics-informed machine learning could be transferred to fairness problems, as described in our recent \textit{arXiv} pre-print~\cite{scher2023conceptual}.

\subsubsection{Reproducibility Aspects of this Habilitation}

I want to highlight the importance of reproducibility for the research field of recommender systems~\cite{beel2016towards,ferrari2021troubling}. This habilitation has provided several resources to foster the reproducibility of the presented research results (see Section~\ref{sec:cont_summary}). In the future, I want to further contribute to the reproducibility of machine learning research in general, and recommender systems research in particular, by discussing barriers and best practices as outlined in our recent publications~\cite{semmelrock2023reproducibility,hicss_repro_2024}. 

\vspace{3mm}
Finally, I hope that the scientific results and findings of this habilitation contribute to advancing research on the trustworthiness of recommender systems.

\newpage


\addcontentsline{toc}{chapter}{Bibliography} 
\bibliographystyle{plain}
\bibliography{bibtex}

\newpage








\end{document}